\documentclass{article}%
\usepackage{graphicx}
\usepackage{amsmath}
\usepackage{amsfonts}
\usepackage{amssymb}%
\providecommand{\U}[1]{\protect\rule{.1in}{.1in}}
\newtheorem{theorem}{Theorem}
\newtheorem{acknowledgement}[theorem]{Acknowledgement}

\newtheorem{corollary}[theorem]{Corollary}

\newtheorem{definition}[theorem]{Definition}

\newtheorem{lemma}[theorem]{Lemma}

\newtheorem{proposition}[theorem]{Proposition}
\newtheorem{remark}[theorem]{Remark}

\newenvironment{proof}[1][Proof]{\textbf{#1.} }{\ \rule{0.5em}{0.5em}}

\begin{document}

\title{\textbf{
 Local Anomalies, Local Equivariant Cohomology and the Variational Bicomplex
 }}
\date{}
\author{\textsc{Roberto Ferreiro P\'{e}rez}
\\Departamento de Econom\'{\i}a Financiera y Contabilidad I
\\Facultad de Ciencias Econ\'omicas y Empresariales, UCM
\\Campus de Somosaguas, 28223-Pozuelo de Alarc\'on, Spain
\\\emph{E-mail:} \texttt{roferreiro@ccee.ucm.es}}
\maketitle

\part{\textbf{Local Cohomology and the Variational Bicomplex}}
\begin{abstract}
The differential forms on the jet bundle $J^{\infty}E$ of a bundle
$E\rightarrow M$ over a compact $n$-manifold $M$ of degree greater
than $n$ determine differential forms on the space $\Gamma(E)$ of
sections of $E$. The forms obtained in this way are called local
forms on $\Gamma(E)$, and its cohomology is called the local
cohomology of $\Gamma(E)$. More generally, if a
group $\mathcal{G}$ acts on $E$, we can define the local $\mathcal{G}%
$-invariant cohomology. The local cohomology is computed in terms
of the cohomology of the jet bundle by means of the variational
bicomplex theory. A similar result is obtained for the local
$\mathcal{G}$-invariant cohomology. Using these results and the
techniques for the computation of the cohomology of invariant
variational bicomplexes in terms of relative Gelfand-Fuchs
cohomology introduced in \cite{AndPoh}, we construct non trivial
local cohomology classes in the important cases of Riemannian
metrics with the action of diffeomorphisms, and connections on a
principal bundle with the action of automorphisms.

\end{abstract}


\smallskip


\section{Introduction\label{introduction}}

Let us recall some basic constructions on the jet bundle geometrical approach
to the variational calculus. Let $p\colon E\rightarrow M$ be a bundle over a
compact, oriented $n$-manifold $M$ without boundary and let $J^{\infty}E$
denote its $\infty$-jet bundle. If $\lambda\in\Omega^{n}(J^{\infty}E)$ is a
lagrangian density, it determines a function $\mathcal{A}$ (the action
functional) in the space $\Gamma(E)$ of sections of $E$ by setting
$\mathcal{A}(s)=\int_{M} (j^{\infty} s)^{*}\lambda$. The exterior differential
$d\mathcal{A}$ of $\mathcal{A}$ is determined in the following way. Let
$s_{t}\in\Gamma(E)$ be a $1$-parameter family of sections of $E$ with
$s=s_{0}$ and let $X\in T_{s}\Gamma(E) \cong\Gamma(M,s^{\ast}V(E))$ be the
vertical vector field along $s$ defined by $X(p)=\left. \frac{ds_{t}(p)}%
{dt}\right\vert _{t=0}$. Then we have
\[
d\mathcal{A}_{s}(X)\!=\! \left.  \frac{d\mathcal{A}(s_{t})}{dt}\right\vert
_{t=0} \!\!\!=\!\!\int_{M}\!\!\!\!\left.  \frac{d(j^{\infty}s_{t})^{*}\lambda
}{dt}\right\vert _{t=0} \!\!\!=\!\!\int_{M}\!\!(j^{\infty}s)^{*}%
(L_{\mathrm{pr}X}\lambda) \!=\!\!\int_{M}\!\!(j^{\infty}s)^{*}(\iota
_{\mathrm{pr}X}d\lambda).
\]
We see that to the form $\lambda\in\Omega^{n}(J^{\infty}E)$ it corresponds the
function $\mathcal{A}\in\Omega^{0}(\Gamma(E))$, whereas to the form
$d\lambda\in\Omega^{n+1}(J^{\infty}E)$ it corresponds the $1$-form
$d\mathcal{A}\in\Omega^{1}(\Gamma(E))$. Generalizing this idea, we have
defined in \cite{equiconn} an integration map $\Im\colon\Omega^{n+k}%
(J^{\infty}E)\rightarrow\Omega^{k}(\Gamma(E))$, by setting
\[
\Im\lbrack\alpha]_{s}(X_{1},\ldots,X_{k})= \int_{M}(j^{\infty}s)^{\ast}
(\iota_{\mathrm{pr}X_{k}}\ldots\iota_{\mathrm{pr}X_{1}}\alpha).
\]
for $\alpha\in\Omega^{n+k}(J^{\infty}E)$ and $X_{1},\ldots,X_{k}\in
T_{s}\Gamma(E)$. Then we have $\Im[\lambda]=\mathcal{A}$ and $\Im
[d\lambda]=d\mathcal{A}=d\Im[\lambda]$. We prove that we have $\Im
[d\alpha]=d\Im[\alpha]$ for any form $\alpha\in\Omega^{n+k}(J^{\infty}E)$. The
forms of the type $\Im[\alpha]$ for $\alpha\in\Omega^{n+k}(J^{\infty}E)$ are
called the local $k$-forms $\Omega_{\mathrm{loc}}^{k}(\Gamma(E))$ on
$\Gamma(E)$, and its cohomology $H_{\mathrm{loc}}^{k}(\Gamma(E))$ the local
cohomology of $\Gamma(E)$. Hence the local forms are those forms on
$\Gamma(E)$ obtained by integration over $M$ of a form on the jet bundle (i.e.
a form depending on a section and its derivatives). For $k=0$ this notion of
locality corresponds precisely to the notion of ``local functional" needed in
quantum field theory for the study of anomaly cancellation. Moreover, in
\cite{anomalies} we show that the anomaly cancellation can be understood in
terms of local cohomology of forms of degree $2$, and we use some of the
results obtained in the present paper solve the problem proposed in
\cite{singer} consisting in explaining the topological nature of local anomalies.

The local cohomology can be studied in terms of the jet bundle by means of the
variational bicomplex theory. Coming back to the example of the variational
calculus, the Euler-Lagrange form $\mathcal{E}(\lambda)\in\Omega
^{n+1}(J^{\infty}E)$ of $\lambda$ satisfies $\Im[\mathcal{E}(\lambda
)]=\Im[d\lambda]=d\mathcal{A}$, and $d\mathcal{A}=0$ if and only if
$\mathcal{E}(\lambda)=0$. Note that both $\mathcal{E}(\lambda)=0$ and
$d\lambda$ determine the same local form $d\mathcal{A}\in\Omega_{\mathrm{loc}%
}^{1}(\Gamma(E))$. However, only $\mathcal{E}(\lambda)$ determines uniquely
the properties of $d\mathcal{A}$. We recall that from the point of view of the
variational bicomplex theory the Euler-Lagrange operator is given by
$\mathcal{E}(\lambda)=I(d_{H}\lambda)$ where $I$ is the interior Euler
operator and $d_{H}$ the horizontal differential. In general, we show in
Section \ref{LC-VB} that given $\alpha\in\Omega^{n+k}(J^{\infty}E)$ with
$k>0$, there are an infinite number of forms on $J^{\infty}E$ determining the
same local form $\Im[\alpha]$ on $\Gamma(E)$, but the interior Euler operator
selects a canonical representative for it $I(\alpha_{n,k})$, that satisfies
$\Im[\alpha]=\Im[I(\alpha_{n,k})]$, and $\Im[\alpha]=0$ if and only if
$I(\alpha_{n,k})=0$. Hence, if we denote by $\mathcal{F}^{k}(J^{\infty
}E)=I(\Omega^{n,k}(J^{\infty}E))$ the space of functional forms, we have the
isomorphisms $\Omega_{\mathrm{loc}}^{k}(\Gamma(E)) \cong\mathcal{F}%
^{k}(J^{\infty}E)$, and $H_{\mathrm{loc}}^{k}(\Gamma(E))\cong H^{k}%
(\mathcal{F}^{\bullet}(J^{\infty}E))\cong H^{n+k}(J^{\infty}E)\cong
H^{n+k}(E)$ for $k>0$.

If a group $\mathcal{G}$ acts on $E$ by automorphisms, we can consider the
local $\mathcal{G}$-invariant cohomology of $\Gamma(E)$, $H_{\mathrm{loc}}%
^{k}(\Gamma(E))^{\mathcal{G}}$, and clearly we also have the isomorphisms
$\Omega_{\mathrm{loc}}^{k}(\Gamma(E))^{\mathcal{G}} \cong\mathcal{F}%
^{k}(J^{\infty}E)^{\mathcal{G}}$, and $H_{\mathrm{loc}}^{k}(\Gamma(E))\cong
H^{k}(\mathcal{F}^{\bullet}(J^{\infty}E))$ for $k>1$. Under certain conditions
analyzed in \cite{AndPoh2,AndPoh} the invariant cohomology of the
Euler-Lagrange complex is isomorphic to the invariant cohomology of
$J^{\infty}E$, and in that case we have $H_{\mathrm{loc}}^{k}(\Gamma
(E))^{\mathcal{G}} \cong H^{n+k}(J^{\infty}E)^{\mathcal{G}}$ for $k>1$.
Moreover, in \cite{AndPoh} it is shown that in certain cases the invariant
cohomology of $J^{\infty}E$ can be computed in terms of relative Lie algebra
cohomology of formal vector fields.

Finally, we apply the preceding constructions to study the local invariant
cohomology of the space of Riemannian metrics $\mathfrak{Met}M$ with the
action of the group $\mathrm{Diff}M$ of diffeomorphisms and the space of
connections $\mathcal{A}_{P}$ on a principal bundle $P$ with the action of the
group $\mathrm{Aut}P$ of automorphisms of $P$. We construct some local
invariant cohomology classes on these spaces and we prove the non triviality
of that classes, by relating them to relative Gelfand-Fuchs cohomology of
formal vector fields. These results are used in \cite{anomalies} to study the
problem of local anomaly cancellation for gravitational and mixed anomalies.

\section{Local cohomology}

\label{LC}

Let $p\colon E\rightarrow M$ be a bundle over a compact, oriented $n$-manifold
$M$ without boundary. We denote by $V(E)$ the vertical bundle and by $J^{r}E$
its $r$-jet bundle. We have the projections $p_{r}\colon J^{r}E\rightarrow M$,
$p_{r,s}\colon J^{r}E\rightarrow J^{s}E$ for $s<r$. Let $\Gamma(E)$ be the
space of global sections of $E$, that we assume to be not empty. $\Gamma(E)$
admits an structure of a Frechet manifold (see \cite[Section I.4]{hamilton}
for details), and for any $s\in$ $\Gamma(E)$, the tangent space to the
manifold $\Gamma(E)$ is isomorphic to the space of vertical vector fields
along $s$, that is $T_{s}\Gamma(E)\simeq\Gamma(M,s^{\ast}V(E)).
\label{tangent}$

A diffeomorphism $\phi\in\mathrm{Diff}E$ is said to be projectable if there
exists $\underline{\phi}\in\mathrm{Diff}M$ satisfying $\phi\circ
p=p\circ\underline{\phi}$. We denote by $\mathrm{Proj}E$ the space of
projectable diffeomorphism of $E$, and we denote by $\mathrm{Proj}^{+}E$ the
subgroup of elements such that $\underline{\phi}\in\mathrm{Diff}^{+}M$,
\emph{i.e.} $\underline{\phi}$ is orientation preserving. The space of
projectable vector fields on $E$ is denoted by $\mathrm{proj}E$, and can be
considered as the Lie algebra of $\mathrm{Proj}E$. We denote by $\phi^{(r)}$
(resp. $\mathrm{pr^{r}}X$) the prolongation of $\phi\in\mathrm{Proj}E$ (resp.
$X\in\mathrm{proj}E$) to $J^{r}E$. The group $\mathrm{Proj}E$ acts naturally
on $\Gamma(E)$ in the following way. If $\phi\in\mathrm{Proj}E$, we define
$\phi_{\Gamma(E)}\in\mathrm{Diff}\Gamma(E)$ by $\phi_{\Gamma(E)}(s)= \phi\circ
s\circ\underline{\phi}^{-1}$, for all $s\in\Gamma(E)$. In a similar way, a
projectable vector field $X\in\mathrm{proj}E$ induces a vector field
$X_{\Gamma(E)}\in\mathfrak{X}(\Gamma(E))$.

Let $\mathrm{j}^{r}\colon M\times\Gamma(E)\rightarrow J^{r}E$, $\mathrm{j}%
^{r}(x,s)=j_{x}^{r}s$ be the evaluation map. We define a map $\Im^{r}%
\colon\Omega^{n+k}(J^{r}E) \longrightarrow\Omega^{k}(\Gamma(E))$, by setting
$\Im^{r}[\alpha]= \int_{M}\left( \mathrm{j}^{r}\right) ^{\ast} \alpha$, for
$\alpha\in\Omega^{n+k}(J^{r}E)$. If $\alpha\in\Omega^{k}(J^{r}E)$ with $k<n$,
we set $\Im^{r}[\alpha]=0$. The operator $\Im$ satisfies the following
properties (see \cite{equiconn})

\begin{proposition}
\label{propF} For any $\alpha\in\Omega^{n+k}(J^{r}E)$ we have

\begin{enumerate}
\item $\Im^{r}[d\alpha]=d\Im^{r}[\alpha]$.

\item $\Im^{r}[(\phi^{(r)})^{\ast}\alpha]= \phi_{\Gamma(E)}^{\ast}\Im
^{r}[\alpha]$, for any $\phi\in\mathrm{Proj}^{+}E$.

\item $\Im^{r}[\iota_{\mathrm{pr}^{r} X}\alpha]= \iota_{X_{\Gamma(E)}}\Im
^{r}[\alpha]$ for any $X\in\mathrm{proj}E$.
\end{enumerate}
\end{proposition}

\begin{corollary}
\label{expresion}Let $\alpha\in\Omega^{n+k}(J^{r}E)$ and $X_{1},\ldots
,X_{k}\in T_{s}\Gamma(E)$. Then we have
\[
\Im^{r} [\alpha]_{s}(X_{1},\ldots,X_{k})=\int_{M}(j^{r}s)^{\ast}
(\iota_{\mathrm{pr}^{r} X_{k}}\ldots\iota_{\mathrm{pr}^{r} X_{1}}\alpha).
\]

\end{corollary}

The forms of the type $\Im^{r}[\alpha]$ for certain $r\in\mathbb{N}$ and
$\alpha\in\Omega^{n+k}(J^{r} E)$ are called local $k$-forms, and the space of
local $k$-forms on $\Gamma(E)$ is denoted by $\Omega_{\mathrm{loc}}^{k}%
(\Gamma(E))$. By Proposition \ref{propF}, $\Omega_{\mathrm{loc}}^{\bullet
}(\Gamma(E))$ is closed under $d$, and we denote by $H_{\mathrm{loc}}%
^{\bullet}(\Gamma(E))$ the cohomology of the complex $(\Omega_{\mathrm{loc}%
}^{\bullet}(\Gamma(E)),d)$. We have an induced map in cohomology
$H_{\mathrm{loc}}^{\bullet}(\Gamma(E)) \rightarrow H^{\bullet}(\Gamma(E))$.
The key point is that this map is not injective in general (e.g. see
\cite{proce} for an example). If $\alpha\in\Omega_{\mathrm{loc}}^{k}%
(\Gamma(E))$ is closed, its cohomology class on $H^{k}(\Gamma(E))$ vanishes if
and only if $\alpha$ is the exterior differential of a form $\beta\in
\Omega^{k-1}(\Gamma(E))$, while its cohomology class on $H_{\mathrm{loc}}%
^{k}(\Gamma(E))$ vanishes if and only if $\alpha$ is the exterior differential
of a \emph{local }form $\beta\in\Omega_{\mathrm{loc}}^{k-1}(\Gamma(E))$.

\section{The variational bicomplex}

\label{VarBi}

We denote by $J^{\infty}E$ the infinite jet bundle (see
\cite{VB,olver,saunders} for the details on the geometry of $J^{\infty}E$). We
have the projections $p_{\infty}\colon J^{\infty}E\rightarrow M$,
$p_{\infty,r}\colon J^{\infty}E\rightarrow J^{r}E$, and $\Omega^{k}(J^{\infty
}E)= \underrightarrow{\mathrm{lim}}\Omega^{k}(J^{r}E)$. We denote by
$\phi^{(\infty)}$ (resp. $\mathrm{pr}X$) the prolongation of $\phi
\in\mathrm{Proj}E$ (resp. $X\in\mathrm{proj}E$) to $J^{\infty}E$.

A local trivialization $(U;x^{i},y^{\alpha})$ of $E$ induces a local
coordinate system $((p_{\infty,0})^{-1}U;x^{i},y^{\alpha},y_{J}^{\alpha})$,
$i=1,\ldots,n$, $j=1,\ldots,m$, $J\in\mathbb{N}^{k}$, $J$ symmetric
$k=1,2,\ldots$ on $J^{\infty}E$, by setting $y_{J}^{\alpha}(j_{x}^{\infty
}s)=\frac{\partial^{\left|  J\right|  } (y^{\alpha}\circ s)}{\partial x^{J}%
}(x)$ for every local section $s$ of $p\colon E\rightarrow M$. If in local
coordinates $X=f^{i}\frac{\partial}{\partial x^{i}}+g^{\alpha} \frac{\partial
}{\partial y^{\alpha}}$, then we have (\cite[Theorem 2.36]{olver})
\begin{equation}
\label{prX}\mathrm{pr}X=f^{i}\frac{\partial}{\partial x^{i}} +g^{\alpha}%
\frac{\partial}{\partial y^{\alpha}}+ \sum\frac{d^{\left|  J\right|  }}%
{dx^{J}} \left(  g^{\alpha}-\sum f^{i} y^{\alpha}_{J+i} \right)  +\sum f^{i}
y^{\alpha}_{J+i}%
\end{equation}
where $\frac{d^{\left|  J\right|  }}{dx^{J}}=\frac{d}{dx^{j_{i}}}\cdots
\frac{d}{dx^{j_{k}}}$, and $\frac{d}{dx^{r}}=\frac{\partial}{\partial x^{r}}
+\sum_{\left|  K\right|  =1}^{\infty}y_{K+r}^{j}\frac{\partial}{\partial
y_{K}^{j}}$.

The evolutionary vector fields are defined as the vertical fields on $E$ with
coefficients in $J^{\infty}E$, i.e, $\mathrm{Ev}(E)=\Gamma(J^{\infty}E,V(E))$.
If $X\in\mathfrak{(}X)(E)$ is a projectable vector field, the evolutionary
part of $X$ is given by $(\mathrm{ev}X)(j_{x}^{\infty}s)=X(s(x))-s_{*}%
(\underline{X}(x))$. The total vector fields are the vector fields on $M$ with
coefficients in $J^{\infty}E$, i.e, $\mathrm{Tot}(E)=\Gamma(J^{\infty}E,TM)$.
Given a projectable vector field $X\in\mathfrak{(}X)(E)$, we have
$\mathrm{pr}X=\mathrm{pr}(\mathrm{ev}X)+\mathrm{tot}X$, where $\mathrm{tot}X$
denotes the total part of $X$. In local coordinates, if $X=f^{i}\frac
{\partial}{\partial x^{i}}+g^{\alpha} \frac{\partial}{\partial y^{\alpha}}$
then we have $\mathrm{tot}X=f^{i}d/dx^{i}$ and $\mathrm{ev}X= (g^{\alpha
}-y^{\alpha}_{i} f^{i})\partial/ \partial y^{\alpha}$.

We define $\gamma\colon\Gamma(E)\times\mathrm{Ev}(E)\rightarrow T\Gamma(E)$ by
setting $\gamma(s,X)=(s,X\circ j^{\infty}s)$. We also use the notation
$\gamma_{s}(X)=\gamma(s,X)$. The following Proposition shows that the vector
field $X_{\Gamma(E)}$ is determined by the evolutionary part of $X$.

\begin{proposition}
If $X\in\mathrm{proj}E$ is a projectable vector field then $X_{\Gamma
(E)}(s)=\gamma_{s}(\mathrm{ev}X)$. In particular, if $X$ is a vertical vector
field then $X_{\Gamma(E)}(s)=X\circ s$
\end{proposition}

\begin{proof}
Let $X\in\mathrm{proj}E$ be a projectable vector field with projection
$\underline{X}$, and let $\Phi_{t}\in\mathrm{Proj}E$ be its flux and
$\underline{\Phi}_{t}\in\mathrm{Diff}M$ its projection onto $M$. Given
$s\in\Gamma(E)$ we have by definition $X_{\Gamma(E)}(s)=\dot{s}_{0}$, where
$s_{t}=\Phi_{t}\circ s\circ\underline{\Phi}_{-t}$ and $\dot{s}_{0}%
=\frac{ds_{t}}{dt}|_{t=0}$. For every $x\in M$ we have $X_{\Gamma
(E)}(s)(x)=\dot{s}_{0}(x)= \dot{\Phi}_{0}(s(x))+D_{s(x)}\Phi_{0}\circ D_{x}
s(-\underline{\dot\Phi}_{0})=X(s(x))-s_{*}(\underline{X}(x)) =(\mathrm{ev}%
X)(j_{x}^{\infty}s)$, where we have used that $\Phi_{0}$ is the identity.
\end{proof}

Let us recall the basic definitions of the variational bicomplex theory (see
\cite{VB,olver} for details). On $J^{\infty}E\rightarrow M$ we have a
bigraduation $\Omega^{k}(J^{\infty}E)=\bigoplus_{k=p+q}\Omega^{p,q}(J^{\infty
}E)$ into horizontal and contact (or vertical) degree. If $\alpha\in\Omega
^{k}(J^{\infty}E)$ we denote by $\alpha_{p,q}\in\Omega^{p,q}(J^{\infty}E)$ its
$p$-horizontal and $q$-contact component. We denote $\Omega^{p,q}(J^{\infty
}E)$ simply by $\Omega^{p,q}$ when there is no risk of confusion. According to
the preceding bigraduation we have a decomposition of the exterior
differential $d=d_{H}+d_{V}$.

We denote by $I\colon\Omega^{n,k}\rightarrow\Omega^{n,k}$ the interior Euler
operator. We recall that it satisfies the following properties: $I^{2}=I$,
$\mathrm{ker}I=d_{H}(\Omega^{n-1,k})$, $Id_{V}=d_{V}I$. The image of the
interior Euler operator $\mathcal{F}^{k}=I(\Omega^{n,k})$ is called the space
of functional $k$-forms. We have $\Omega^{n,k}\cong\mathcal{F}^{k}\oplus
d_{H}(\Omega^{n-1,k})$, i.e. $\mathcal{F}^{k}\cong\Omega^{n,k}/d_{H}%
(\Omega^{n-1,k})$.

The vertical differential $d_{V}$ induces a differential in the space of
functional forms $\delta_{V}\colon\mathcal{F}^{k}\rightarrow\mathcal{F}^{k+1}%
$, $\delta_{V}\alpha=I(d_{V}\alpha)$. We have the usual diagram for the
augmented variational bicomplex%

\[%
\begin{array}
[c]{ccccccccccc}%
\uparrow{\scriptstyle d_{V}} &  &  &  &  &  & \uparrow{\scriptstyle d_{V}} &
& \uparrow{\scriptstyle d_{V}} &  & \uparrow{\scriptstyle\delta_{V}}\\
\Omega^{0,3} &  &  &  &  &  & \Omega^{n-1,3} & \overset{d_{H}}{\longrightarrow
} & \Omega^{n,3} & \!\!\!\!\overset{I}{\longrightarrow}\!\!\!\! &
\mathcal{F}^{3}\\
\uparrow{\scriptstyle d_{V}} &  & \uparrow{\scriptstyle d_{V}} &  &  &  &
\uparrow{\scriptstyle d_{V}} &  & \uparrow{\scriptstyle d_{V}} &  &
\uparrow{\scriptstyle\delta_{V}}\\
\Omega^{0,2} & \overset{d_{H}}{\longrightarrow} & \Omega^{1,2} &
\overset{d_{H}}{\longrightarrow} & \ldots & \overset{d_{H}}{\longrightarrow} &
\Omega^{n-1,2} & \overset{d_{H}}{\longrightarrow} & \Omega^{n,2} &
\!\!\!\!\overset{I}{\longrightarrow}\!\!\!\! & \mathcal{F}^{2}\\
\uparrow{\scriptstyle d_{V}} &  & \uparrow{\scriptstyle d_{V}} &  &  &  &
\uparrow{\scriptstyle d_{V}} &  & \uparrow{\scriptstyle d_{V}} &  &
\uparrow{\scriptstyle\delta_{V}}\\
\Omega^{0,1} & \overset{d_{H}}{\longrightarrow} & \Omega^{1,1} &
\overset{d_{H}}{\longrightarrow} & \ldots & \overset{d_{H}}{\longrightarrow} &
\Omega^{n-1,1} & \overset{d_{H}}{\longrightarrow} & \Omega^{n,1} &
\!\!\!\!\overset{I}{\longrightarrow}\!\!\!\! & \mathcal{F}^{1}\\
\uparrow{\scriptstyle d_{V}} &  & \uparrow{\scriptstyle d_{V}} &  &  &  &
\uparrow{\scriptstyle d_{V}} &  & \uparrow{\scriptstyle d_{V}} & \quad
\nearrow{\scriptstyle\delta_{V}}\!\!\!\! & \\
\Omega^{0,0} & \overset{d_{H}}{\longrightarrow} & \Omega^{1,0} &
\overset{d_{H}}{\longrightarrow} & \ldots & \overset{d_{H}}{\longrightarrow} &
\Omega^{n-1,0} & \overset{d_{H}}{\longrightarrow} & \Omega^{n,0} &  &
\end{array}
\]

The Euler-Lagrange complex $\mathcal{E}^{\bullet}(J^{\infty}E)$ is the
following complex
\[
\Omega^{0,0}\overset{d_{H}}{\longrightarrow} \Omega^{1,0}\overset{d_{H}%
}{\longrightarrow} \ldots\overset{d_{H}}{\longrightarrow}\Omega^{n-1,0}
\overset{d_{H}}{\longrightarrow}\Omega^{n,0} \overset{{\delta_{V}}%
}{\longrightarrow}\mathcal{F}^{1} \overset{{\delta_{V}}}{\longrightarrow}
\mathcal{F}^{2}\overset{{\delta_{V}}}{\longrightarrow} \cdots
\]

We recall that in the jet bundle formulation of the variational calculus a
Lagrangian density is an element $\lambda\in\Omega^{n,0}$, the map $\delta
_{V}\colon\Omega^{n,0}\rightarrow\mathcal{F}^{1}$ is the Euler-Lagrange map
(i.e. $\delta_{V}\lambda$ is the Euler-Lagrange operator of $\lambda$), and
the map $\delta_{V}\colon\mathcal{F}^{1}\rightarrow\mathcal{F}^{2}$ is the
Helmholtz-Sonin mapping characterizing locally variational operators. A
classical result in the variational bicomplex theory (see \emph{e.g.}
\cite{VB,takens}) asserts that $H^{\bullet}(\mathcal{E}^{\bullet}(J^{\infty
}E)) \cong H^{\bullet}(J^{\infty}E)\cong H^{\bullet}(E)$. This result is based
on the fact that the interior rows of the variational bicomplex are exact
\[%
\begin{array}
[c]{ccccccccccccc}%
0 & \!\rightarrow\! & \Omega^{0,k} & \!\overset{d_{H}}{\longrightarrow}\! &
\Omega^{1,k} & \!\overset{d_{H}}{\longrightarrow}\! & \ldots & \!\overset
{d_{H}}{\longrightarrow}\! & \Omega^{n,k} & \!\overset{I}{\longrightarrow}\! &
\mathcal{F}^{k} & \!\rightarrow\! & 0.
\end{array}
\label{introws}%
\]

\section{Local forms and the variational bicomplex}

\label{LC-VB}

The family of maps ${\Im^{r}}$ determine a map $\Im\colon\Omega^{n+k}%
(J^{\infty}(E))\rightarrow\Omega^{k}(\Gamma(E))$, and by definition we have
$\Omega^{k}_{\mathrm{loc}}(\Gamma(E))=\Im(\Omega^{n+k}(J^{\infty}(E)))$. We
now study the relation between the map $\Im$ and the variational bicomplex.

\begin{proposition}
For $\alpha\in\Omega^{n+k}(J^{\infty}E)$, $k>0$, we have $\Im[\alpha
]=\Im[\alpha_{n,k}]=\Im[I(\alpha_{n,k})]$.
\end{proposition}

\begin{proof}
That $\Im[\alpha]=\Im[\alpha_{n,k}]$ follows from Proposition \ref{expresion}.
As we have $\alpha_{n,k}=I(\alpha_{n,k})+d_{H} \eta$ for certain $\eta
\in\Omega^{n-1,k}$, to prove that $\Im[\alpha_{n,k}]=\Im[I(\alpha_{n,k})]$ it
is sufficient to prove that $\Im[d_{H} \eta]=0$. As $d_{V} \eta\in
\Omega^{n-1,k+1}$ we have $\Im[d_{V} \eta]=0$. Hence $\Im[d_{H} \eta
]=\Im[d\eta]=d\Im[\eta]=0$, where we have used that $\Im[\eta]=0$ because
$\eta\in\Omega^{n-1,k}$.
\end{proof}

In \cite{VB} (see also \cite[section 5.4]{olver}) another more general
interpretation of the functional forms is given. Precisely, every form
$\alpha\in\Omega^{n+k}(J^{\infty}(E))$ determines a multilineal map on
evolutionary vector fields $\mathcal{W}[\alpha]\colon\Gamma(E)\times
\bigwedge^{k}\mathrm{Ev}(E) \rightarrow\mathbb{R}$, $\mathcal{W}[\alpha
]_{s}(X_{1},\dots,X_{k})=\int_{M}(j^{\infty}s)^{\ast} (\iota_{\mathrm{pr}
X_{k}}\ldots\iota_{\mathrm{pr}X_{1}}\alpha)$ for $s\in\Gamma(E)$ and
$X_{1},\dots,X_{k}\in\mathrm{Ev}(E)$. The relation between $\Im$ and
$\mathcal{W}$ is that we have  $\mathcal{W}[\alpha]_{s}(X_{1},\dots,X_{k})
=\Im[\alpha]_{s}(\gamma_{s}(X_{1}),\dots,\gamma_{s}(X_{k})) $. Conversely, if
$X_{1},\dots,X_{k}\in T_{s}\Gamma(E)\cong\Gamma(s^{*}V(E))$ and $\bar{X}%
_{1},\dots,\bar{X}_{k}$ are vertical vector fields on $E$ extending
$X_{1},\dots,X_{k}$ then we have $\Im[\alpha]_{s}(X_{1},\dots,X_{k})=
\mathcal{W}[\alpha]_{s}(\bar{X}_{1},\dots,\bar{X}_{k})$. Hence $\mathcal{W}%
[\alpha]$ is completely determined by $\Im[\alpha]$ and vice versa.

In \cite[Proposition 3.1]{VB} (see also \cite[Lemma 5.85]{olver}) it is proved
that $\mathcal{W}[\alpha]=0$ if and only if $I( \alpha_{n,k})=0$. By the
preceding considerations we have $\mathcal{W}[\alpha]=0$ if and only if
$\Im[\alpha]=0$. Hence we have the following

\begin{theorem}
\label{injec} For $\alpha\in\Omega^{n+k}(J^{\infty}E)$, $k>0$, we have
$\Im[\alpha]=0$ if and only if $I( \alpha_{n,k})=0$.
\end{theorem}

Hence the map $\Im$ is uniquely determined by its restriction to the space of
functional forms $\mathcal{F}^{k}(J^{\infty}E)$\ and this restriction
$\mathcal{F}^{k}(J^{\infty}E)\hookrightarrow\Omega^{k}(\Gamma(E))$ is injective.

\begin{corollary}
\label{cor}For $\alpha\in\Omega^{n+k}(J^{\infty}E)$, $k>0$, we have
$d\Im\lbrack\alpha]=\Im\left[  \delta_{V}I(\alpha_{n,k})\right]  $, and
$d(\Im\lbrack\alpha])=0$ if and only if $\delta_{V}I(\alpha_{n,k})=0$.
\end{corollary}

\begin{corollary}
For every $k\geq1$ the integration map $\Im$ induces isomorphisms
$\mathcal{F}^{k}(J^{\infty}E) \cong\Omega_{\mathrm{loc}}^{k}(\Gamma(E))$, and
$H_{\mathrm{loc}}^{k}(\Gamma(E)) \cong H^{n+k}(\mathcal{E}^{\bullet}%
(J^{\infty}E)) \cong H^{n+k}(E)$.
\end{corollary}

Let us analyze now what happens for $k=0$. In this case we do not have an
interior Euler operator, and hence we can not select a canonical lagrangian
for a given local functional. We can define $\mathcal{F}^{0}(J^{\infty
}E)=\Omega^{n,0}/d_{H} \Omega^{n-1,0}$, and we have a map $\Im\colon
\mathcal{F}^{0}(J^{\infty}E)\rightarrow\Omega^{0}(\Gamma(E))$. However, this
map is not injective. For example (see \cite{VB}) we can consider the bundle
$E=S^{2}\times T^{2}\rightarrow S^{2}$, and $\alpha,\beta\in\Omega^{1}(T^{2})$
generators of $H^{1}(T^{2})$. If we take $\omega=\alpha\wedge\beta\in
\Omega^{2}(E)$, then $0\neq[\omega]\in H^{2}(E)$ and hence $\omega$ defines a
non trivial element in $\mathcal{F}^{0}(J^{\infty}E)$. However we have
$\Im[\omega]=0$, as for every $s\in\Gamma(E)$ we have $\Im[\omega]_{s}%
=\int_{S^{2}}(j^{\infty}s)^{*}\omega=\int_{S^{2}}(j^{\infty}s)^{*}\alpha
\wedge(j^{\infty}s)^{*}\beta=0$, where the last equality follows as we have
$(j^{\infty}s)^{*}\alpha=d\eta$ for certain $\eta\in\Omega^{1}(S^{2})$ because
$H^{1}(S^{2})=0$.

Let $\ker(\Im_{0})\subset\Omega^{n,0}$ denote the space of trivial lagrangian
densities. Clearly we have by definition $\Omega^{0}_{\mathrm{loc}}(\Gamma(E))
\cong\Omega^{n,0}/\ker{\Im_{0}}$. Moreover, if $\delta_{V}^{0}$ denotes the
Euler-Lagrange operator $\delta_{V}^{0}\colon\Omega^{n,0}\to\mathcal{F}^{1}$,
we have $d_{H}(\Omega^{n-1,0}) \subset\ker(\Im_{0}) \subset\ker(\delta_{V}%
^{0})$. If we define $\mathcal{N}=\ker{\Im_{0}}/d_{H}(\Omega^{n-1,0}) $ we
have a natural inclusion $\mathcal{N}\subset H^{n}(\mathcal{E}^{\bullet
}(J^{\infty}E)\cong H^{n} (E)$, and we have $H^{0}_{\mathrm{loc}}(\Gamma(E))
\cong\ker(\delta^{0}_{V})/\ker{\Im_{0}} \cong H^{n}(E)/\mathcal{N}$. In this
way we can identify $\mathcal{N}\!\cong\!\bigcap_{s\in\Gamma(E)} \ker
\{s^{*}\colon\! H^{n}(E)\!\rightarrow\!H^{n}(M)\!\cong\mathbb{R}\}$. If
$\mathcal{N}$ vanishes, we have $\Omega^{0}_{\mathrm{loc}}(\Gamma(E))
\cong\Omega^{n,0}/d_{H} (\Omega^{n-1,0})$ and $H^{0}_{\mathrm{loc}}(\Gamma(E))
\cong H^{n}(E)$. This happens for example if $H^{n}(E)\cong\mathbb{R}$, as it
is easily seen.

We can consider this computation of $H^{0}_{\mathrm{loc}}(\Gamma(E))$ as a
refinement of the inverse problem of the calculus of variations, that is, the
computation of the variationally trivial lagrangian densities modulo
divergences $\ker(\delta^{0}_{V})/d_{H} (\Omega^{n-1,0})\cong H^{n}(E)$. If in
place of lagrangian densities we consider local functionals, and we ask for
the local functionals which are closed, then what we obtain is $H^{0}%
_{\mathrm{loc}}(\Gamma(E))\!\cong H^{n}(E)/\mathcal{N}$.

\section{Local invariant cohomology \label{VB-GF}}

Let us assume now that a Lie group $\mathcal{G}$ acts on $E\rightarrow M$ by
elements of $\mathrm{Proj}^{+}E$. Then we have an induced action of
$\mathcal{G}$ on $J^{\infty}E$ and the variational bicomplex remains invariant
under this action. By considering $\mathcal{G}$-invariant forms we obtain the
$\mathcal{G}$-invariant variational bicomplex and the $\mathcal{G}$-invariant
Euler-Lagrange complex. We define the space of local $\mathcal{G}$-invariant
forms $\Omega_{\mathrm{loc}}^{k}(\Gamma(E))^{\mathcal{G}}$ as the subspace of
$\mathcal{G}$-invariant elements on $\Omega_{\mathrm{loc}}^{k}(\Gamma(E))$,
and the local $\mathcal{G}$-invariant cohomology as the cohomology of this
complex. As the integration map $\Im$ is $\mathcal{G}$-equivariant, from
Theorem \ref{injec} we obtain the following

\begin{corollary}
\label{cor2}For every $k\geq1$ the integration map $\Im$ induces isomorphisms
$\mathcal{F}^{k}(J^{\infty}E)^{\mathcal{G}} \cong\Omega_{\mathrm{loc}}%
^{k}(\Gamma(E))^{\mathcal{G}}$ and we have $\Omega_{\mathrm{loc}}^{k}%
(\Gamma(E))^{\mathcal{G}} =\Im(\Omega^{n+k}(J^{\infty}E)^{\mathcal{G}})$.
Moreover, $\Im$ induces isomorphisms $H_{\mathrm{loc}}^{k}(\Gamma
(E))^{\mathcal{G}} \cong H^{n+k}(\mathcal{E}^{\bullet} (J^{\infty
}E))^{\mathcal{G}}$ for $k>1$.
\end{corollary}

Let us analyze now what happens for the local invariant cohomology of order
$0$ and $1$. We have $\Omega^{0}_{\mathrm{loc}}(\Gamma(E))^{\mathcal{G}}
\cong(\Omega^{n,0}/\ker(\Im_{0}))^{\mathcal{G}}$, and hence $H^{0}%
_{\mathrm{loc}}(\Gamma(E)) \cong(\ker(\delta_{V}^{0})/\ker(\Im_{0}%
))^{\mathcal{G}} \cong\{H^{n}(E)/\mathcal{N}\}^{\mathcal{G}}$, where
$\{H^{n}(E)/\mathcal{N}\}^{\mathcal{G}}$ denotes the space of $\mathcal{G}%
$-invariant elements in $H^{n}(E)/\mathcal{N}$. In particular, if
$\mathcal{N}=0$ we have $H^{0}_{\mathrm{loc}}(\Gamma(E) \cong\{H^{n}%
(E)\}^{\mathcal{G}}$. If $\mathcal{G}$ is connected we clearly have
$\{H^{n}(E)\}^{\mathcal{G}}\cong H^{n}(E)$ and hence $H^{0}_{\mathrm{loc}%
}(\Gamma(E))^{\mathcal{G}} \cong H^{n}(E) \cong H^{0}_{\mathrm{loc}}%
(\Gamma(E)$. If there are elements of $\mathcal{G}$ not homotopic to the
identity we can have $\{H^{n}(E)\}^{\mathcal{G}}\ncong H^{n}(E)$. For example,
if we consider the bundle $E=S^{1}\times T^{2}\rightarrow S^{1}$, where
$T^{2}=S^{1} \times S^{1}$ is the $2$-torus and $\mathcal{G}=\mathrm{Diff}%
T^{2}$ then we have $H^{0}_{\mathrm{loc}}(\Gamma(E)) \cong H^{1}%
(E)\cong\mathbb{R}^{3}$ and $H^{0}_{\mathrm{loc}}(\Gamma(E)) ^{\mathcal{G}}
\cong\{H^{1}(E)\}^{\mathcal{G}} \cong\mathbb{R}$. The last isomorphism follows
because the cohomology classes of $H^{1}(E)$ coming from $T^{2}$ are not
$\mathrm{Diff}T^{2}$-invariant.

The situation for $H^{1}_{\mathrm{loc}}(\Gamma(E))^{\mathcal{G}}$ is more
complicated. That in general we do not have $H_{\mathrm{loc}}^{1}%
(\Gamma(E))^{\mathcal{G}}\cong H^{n+1}(\mathcal{E}^{\bullet}(J^{\infty}
E))^{\mathcal{G}}$ is shown in Section \ref{metrics} with a concrete example
of gravitational Chern-Simons terms. The difference between $\mathcal{G}%
$-invariant cohomology and ordinary cohomology is that we have $\Omega
^{0}_{\mathrm{loc}}(\Gamma(E))^{\mathcal{G}} \cong(\Omega^{n,0}/\ker(\Im
_{0}))^{\mathcal{G}}$ which in general is not equal to $(\Omega^{n,0}%
)^{\mathcal{G}}/\ker(\Im_{0})$. We say that a lagrangian density $\lambda
\in\Omega^{n,0}$ is weakly $\mathcal{G}$-invariant if for every $\phi
\in\mathcal{G}$ we have $(\phi^{(\infty)})^{*}(\lambda)-\lambda\in\ker{\Im
_{0}}$. At the Lie algebra level we have $L_{\mathrm{pr}X}\lambda\in\ker
{\Im_{0}}$ for every $X\in{\mathrm{Lie}\mathcal{G}}$. If we denote by
$(\Omega^{n,0})^{\mathcal{G}_{W}}$ the space of weakly $\mathcal{G}$-invariant
lagrangians, then we have $\Omega^{0}_{\mathrm{loc}}(\Gamma(E))^{\mathcal{G}}
\cong(\Omega^{n,0})^{\mathcal{G}_{W}}/\ker{\Im_{0}}$. Also note that if
$\lambda\in(\Omega^{n,0})^{\mathcal{G}_{W}}$ then $\delta_{V}^{0}(\lambda
)\in(\mathcal{F}^{1}(J^{\infty}E))^{\mathcal{G}}$ as we have $(\phi^{(\infty
)})^{*}(\delta_{V}^{0}(\lambda)) =\delta_{V}^{0}((\phi^{(\infty)})^{*}%
(\lambda)) =\delta_{V}^{0}(\lambda)$, where we have used that $\delta_{V}^{0}$
vanishes on $\ker\Im_{0}$. Hence we have
\[
H_{\mathrm{loc}}^{1}(\Gamma(E))^{\mathcal{G}} \cong\dfrac{(\ker\delta_{V}%
^{1})^{\mathcal{G}}} {\delta_{V}^{0}((\Omega^{n,0})^{\mathcal{G}_{W}}) }
\cong\dfrac{H^{n+1}(\mathcal{E}^{\bullet}(J^{\infty}(E)))^{\mathcal{G}}}
{[\delta_{V}^{0}((\Omega^{n,0})^{\mathcal{G}_{W}})]}.
\]

Under very general conditions the interior rows of the $\mathcal{G}$-invariant
variational bicomplex are exact. For example this happens if there exists a
$\mathcal{G}$-invariant torsion-free horizontal connection $\nabla$ on the
space $\mathrm{tot}(J^{\infty}E)\cong\Gamma(J^{\infty}E,TM)$ of total vector
fields (see \cite{AndPoh2}). In that case, the homotopy operators used to
establish the exactness of (\ref{introws}) can be modified by substituting the
ordinary derivatives by covariant derivatives with respect to $\nabla$, thus
obtaining $\mathcal{G}$-invariant homotopy operators which can be used to
prove the exactness of the interior rows of the $\mathcal{G}$-invariant
variational bicomplex%

\[%
\begin{array}
[c]{ccccccccccccc}%
0\! & \!\!\rightarrow\!\! & \!\!\left(  \Omega^{0,k}\right)  ^{\mathcal{G}
}\!\! & \!\!\overset{d_{H}}{\longrightarrow}\!\! & \!\!\left(  \Omega
^{1,k}\right)  ^{\mathcal{G}}\!\! & \!\!\overset{d_{H}}{\longrightarrow}\!\! &
\!\ldots\! & \!\!\overset{d_{H}}{\longrightarrow}\!\! & \!\!\left(
\Omega^{n,k}\right)  ^{\mathcal{G}}\!\! & \!\!\overset{I}{\longrightarrow
}\!\! & \!\!\left(  \mathcal{F}^{k}\right)  ^{\mathcal{G}}\!\! &
\!\!\rightarrow\!\! & \!0
\end{array}
\]
for $k>0$. In that case we have isomorphisms for $k>1$
\begin{equation}
H_{\mathrm{loc}}^{k}(\Gamma(E))^{\mathcal{G}}\cong H^{n+k}(\mathcal{E}
^{\bullet}(J^{\infty}E))^{\mathcal{G}}\cong H^{n+k}(J^{\infty}E)^{\mathcal{G}%
}.\label{isocoho}%
\end{equation}

In \cite{AndPoh} it is shown that the invariant cohomology $H^{n+k}(J^{\infty
}E)^{\mathcal{G}}$ of the jet bundle can be determined in certain cases in
terms of relative Lie algebra cohomology of formal vector fields. We apply
this idea in the following sections in order to study the local invariant
cohomology of the spaces of Riemmanian metrics and connections on principal bundles.

\section{Riemannian metrics and diffeomorphisms\label{metrics}}

\subsection{Universal Pontryagin and Euler forms on $J^{1}\mathcal{M}_{M}$}

Let $M$ be a compact and connected $n$-manifold without boundary, and $TM$ its
tangent bundle. We define its bundle of Riemannian metrics $q\colon
\mathcal{M}_{M}\rightarrow M$ by $\mathcal{M}_{M}=\{g_{x}\in S^{2}(T_{x}%
^{\ast}M):g_{x}$ is positive defined on $T_{x}M\}$. Let $\mathfrak{Met}%
M=\Gamma(M,\mathcal{M}_{M})$ denote the space of Riemannian metrics on $M$. We
denote by $\mathrm{Diff}M$ the diffeomorphisms group of $M$, by $\mathrm{Diff}%
^{+}M$ its subgroup of orientation preserving diffeomorphisms and by
$\mathrm{Diff}^{e}M$ the connected component of the identity in $\mathrm{Diff}%
M$. We denote by $q_{1}\colon J^{1}\mathcal{M}_{M}\rightarrow M$ the $1$-jet
bundle of $\mathcal{M}_{M}$ and by $\pi\colon FM\rightarrow M$ the linear
frame bundle of $M$. The pull-back bundle $\bar{q}_{1}\colon q_{1}^{\ast}
FM\rightarrow J^{1}\mathcal{M}_{M}$ is a principal $Gl(n,\mathbb{R})$-bundle
and we have the following commutative diagram
\[%
\begin{array}
[c]{ccc}%
q_{1}^{\ast}FM & \overset{\bar{q}_{1}}{\longrightarrow} & FM\\
{\scriptstyle \bar{\pi}}\downarrow &  & \downarrow{ \scriptstyle\pi}\\
J^{1}\mathcal{M}_{M} & \overset{q_{1}}{\longrightarrow} & M
\end{array}
\]

Every system of coordinates $(U;x^{i})$ on $M$ induces a system of coordinates
$(q^{-1}U;x^{i},y_{ij})$ on $\mathcal{M}_{M}$ by setting $g_{x}=y_{ij}%
(g_{x})(dx^{i})_{x}\otimes(dx^{j})_{x}$, $\forall g_{x}\in\mathcal{M}_{M}$,
$x\in U$. We denote by $(y^{ij})$ the inverse matrix of $(y_{ij})$. Let
$(q_{1}^{-1}U;x^{i},y_{ij,}y_{ij,k})$, be the coordinate system on
$J^{1}\mathcal{M}_{M}$ induced by $(q^{-1}U;x^{i},y_{ij})$; i.e.,
$y_{ij,k}(j_{x}^{1}g)=(\partial(y_{ij}\circ g)/\partial x^{k})(x)$. Note that
if $(U;x^{i})$ is a normal coordinate system for the metric $g$ centered at
$x$, then we have $y_{ij}(j_{x}^{1}g)=\delta_{ij}$, $y_{ij,k}(j_{x}^{1}g)=0$.

The diffeomorphism group of $M$ acts in a natural way on $\mathcal{M}_{M}$. If
$\phi\in\mathrm{Diff}M$, its lift to the bundle of metrics $\bar{\phi}%
\colon\mathcal{M}_{M}\to\mathcal{M}_{M}$ is defined by $\bar{\phi}(g_{x})
=\left(  \phi^{\ast}\right)  ^{-1}(g_{x}) \in(\mathcal{M}_{M})_{\phi(x)}$,
$\phi^{\ast}\colon S^{2}T_{\phi(x)}^{\ast}M \to S^{2}T_{x}^{\ast}M$ being the
induced homomorphism. Hence $q\circ\bar{\phi}=\phi\circ q$. In the same way,
the lift of a vector field $X\in\mathfrak{X}(M)$ is denoted by $\bar{X}%
\in\mathfrak{X}(\mathcal{M}_{M})$. If $X\in\mathfrak{X}(M)$ is given in local
coordinates by $X=X^{i}\partial/\partial x^{i}$, then its lift $\bar{X}%
\in\mathfrak{X}(\mathcal{M}_{M})$ to $\mathcal{M}_{M}$ is given by
\begin{equation}
\bar{X}=X^{i}\frac{\partial}{\partial x^{i}}-\sum_{i\leq j}\left(
\frac{\partial X^{k}}{\partial x^{i}}y_{kj}+\frac{\partial X^{k}}{\partial
x^{j}}y_{ki}\right)  \frac{\partial}{\partial y_{ij}}.\label{expXbarra}%
\end{equation}

We consider the principal $SO(n)$-bundle $O^{+}M\rightarrow J^{1}
\mathcal{M}_{M}$ where $O^{+}M=\left\{  (j_{x}^{1}g,u_{x})\in q_{1}^{\ast
}FM\colon\text{ }u_{x}\,\text{is\emph{ }}g_{x}\text{-orthonormal and
positively oriented}\right\}  $. In \cite{natconn} it is shown that there
exists a connection form $\mbox{\boldmath$\omega$} \in\Omega^{1}%
(O^{+}M,\mathfrak{so}(n))$ (called the universal Levi-Civita connection) on
$O^{+}M$ invariant under the natural action of the group $\mathrm{Diff}^{+}M$.
Let us recall how this connection is constructed. We define a $\mathrm{Diff}%
M$-invariant connection $\mbox{\boldmath$\omega$} _{\mathrm{hor}}\in\Omega
^{1}(q_{1}^{\ast}FM,\mathfrak{gl}(n))$ on $q_{1}^{\ast}FM\rightarrow
J^{1}\mathcal{M}_{M}$ by setting $\mbox{\boldmath$\omega$}_{\mathrm{hor}}(X)
=\omega^{g}((\bar{q}_{1})_{\ast}X)$, for every $X\in T_{(j_{x}^{1}g,u)}%
(q_{1}^{\ast}FM)$, where $\omega^{g}$ denotes the Levi-Civita connection of
the metric $g$. The connection $\mbox{\boldmath$\omega$}_{\mathrm{hor}}$ is
not reducible to the $SO(n)$-bundle $O^{+}M\rightarrow J^{1}\mathcal{M}_{M}$,
but in \cite{natconn} it is shown that it is possible to obtain a reducible
and $\mathrm{Diff}^{+}M$-invariant connection $\mbox{\boldmath$\omega$}$ by
adding to $\mbox{\boldmath$\omega$}_{\mathrm{hor}}$  a contact form $\frac
{1}{2}\mbox{\boldmath$\vartheta$} \in\Omega^{1}(J^{1}\mathcal{M}%
_{M},\mathrm{End}TM)$. We denote by $\mbox{\boldmath$\Omega$}$ and
$\mbox{\boldmath$\Omega$}_{\mathrm{hor}}$ the curvature form of
$\mbox{\boldmath$\omega$}$ and $\mbox{\boldmath$\omega$}_{\mathrm{hor}}$
respectively. In local coordinates, the expressions of
$\mbox{\boldmath$\Omega$}_{\mathrm{hor}}$ and $\mbox{\boldmath$\vartheta$}$
are given by (see \cite{natconn})%

\begin{equation}
\left( \mbox{\boldmath$\Omega$} _{\mathrm{hor}}\right) ^{i}_{j}=
d\mbox{\boldmath$\Gamma$}_{jk}^{i}\wedge dx^{k}+ \mbox{\boldmath$\Gamma$}_{as}%
^{i} \mbox{\boldmath$\Gamma$}_{jr}^{a}dx^{s}\wedge dx^{r} ,\label{Omegahor}%
\end{equation}
\begin{equation}
\mbox{\boldmath$\vartheta$} ^{i}_{j}=y^{ia}(dy_{aj}-y_{aj,k} dx^{k}%
)\label{vartheta}%
\end{equation}
where $\mbox{\boldmath$\Gamma$}_{jk}^{i}= \tfrac{1}{2}y^{ia}(y_{aj,k}%
+y_{ak,j}-y_{jk,a})$.

For any Weil polynomial $p\in I_{r}^{SO(n)}$ the universal characteristic form
$p(\mbox{\boldmath$\Omega$})\in\Omega^{4k}(J^{1}\mathcal{M}_{M})$
corresponding to $p$ is defined as the form obtained by means of the
Chern-Weil theory of characteristic classes by applying $p$ to the curvature
$\mbox{\boldmath$\Omega$}$ of the universal Levi-Civita connection
$\mbox{\boldmath$\omega$}$. In particular we have the universal $k$-th
Pontryagin form of $M$, $p_{k}(\mbox{\boldmath$\Omega$}) \in\Omega^{4k}%
(J^{1}\mathcal{M}_{M})$ and, for $n$ even, the universal Euler form
$\chi(\mbox{\boldmath$\Omega$}) =\frac{1}{(2\pi)^{n}}\mathrm{Pf}%
(\mbox{\boldmath$\Omega$}) \in\Omega^{n}(J^{1}\mathcal{M}_{M})$, where
$\mathrm{Pf}$ denotes the Pfaffian. These forms are closed, $\mathrm{Diff}%
^{+}M$-invariant and satisfy the following universal property (see
\cite{natconn}): for every Riemannian metric $g$ we have $(j^{1}g)^{\ast
}(p(\mbox{\boldmath$\Omega$}))=p(\Omega^{g})$, where $\Omega^{g}\in\Omega
^{2}(M,\mathrm{End}TM)$ is the curvature form of the Levi-Civita connection of
the metric $g$. Hence the Pontryagin forms of degree less or equal than $n$
determine the Pontryagin classes of $M$, while the Pontryagin forms of degree
greater than $n$ determine, by means of $\Im$, closed $\mathrm{Diff}^{+}%
M$-invariant forms on $\mathfrak{Met}M$.

\subsection{Local invariant cohomology of $\mathfrak{Met}M$}

The local cohomology of $\mathfrak{Met}M$ is easily computed. We have
$H_{\mathrm{loc}}^{k}(\mathfrak{Met}M) \cong H^{n+k}(J^{\infty}\mathcal{M}%
_{M}) \cong H^{n+k}(\mathcal{M}_{M}) \cong H^{n+k}(M)=0$ for $k>0$. Moreover,
as $H^{n}(\mathcal{M}_{M})\cong H^{n}(M)=\mathbb{R}$, by the results explained
in Section \ref{LC-VB} we have $H_{\mathrm{loc}}^{0}(\mathfrak{Met}M)
\cong\mathbb{R}$.

Hence $p(\mbox{\boldmath$\Omega$})$ is an exact form for every $p\in
I^{SO(n)}_{r}$ with $2r>n$. In fact it is easy to construct explicitly a form
$\alpha$ satisfying $p(\mbox{\boldmath$\Omega$})=d\alpha$ by fixing a metric
$g_{0}\in\mathfrak{Met}M$. Let $\omega^{g_{0}}$ be the Levi-Civita connection
of $g_{0}$, considered as a connection on the frame bundle $FM$. The
connections $\bar{q}_{1}^{\ast}\omega^{g_{0}}$ and $\mbox{\boldmath$\omega$}$
are both connections on the same bundle $q_{1}^{\ast}FM\rightarrow
J^{1}\mathcal{M}_{M}$, and hence we have $d(Tp(\mbox{\boldmath$\omega$},
\bar{q}_{1}^{\ast}\omega^{g_{0}})) =p(\mbox{\boldmath$\Omega$}) -p(\bar{q}%
_{1}^{\ast}\Omega^{g_{0}})= p(\mbox{\boldmath$\Omega$})$, where
$Tp(\mbox{\boldmath$\omega$}, \bar{q}_{1}^{\ast}\omega^{g_{0}}) \in
\Omega^{n+1}(J^{\infty} \mathcal{M}_{M})^{\mathrm{Diff^{+}}M}$ is the
transgression form corresponding to $\mbox{\boldmath$\omega$}$ and
$\omega^{g_{0}}$ (see \cite{KN}), and we have used that $p(\bar{q}_{1}^{\ast
}\Omega^{g_{0}}) =q_{1}^{\ast}p(\Omega^{g_{0}})=0$ by dimensional reasons.

As the connection $\mbox{\boldmath$\omega$}_{\mathrm{hor}}$ determines a
torsion free connection on the space of total vector fields on $J^{\infty
}\mathcal{M}_{M}$, by the results explained in Section \ref{VB-GF} we have the following

\begin{proposition}
For every $k>1$ the integration map $\Im$ induces isomorphisms
$H_{\mathrm{loc}}^{k} (\mathfrak{Met}M)^{\mathrm{Diff}^{e}M} \cong
H^{n+k}(J^{\infty} \mathcal{M}_{M})^{\mathrm{Diff}^{e}M}$.
\end{proposition}

This result is valid as well for the group $\mathrm{Diff}^{+}M$. In Remark
\ref{contra} we show that this result is not true for $k=1$.

The invariant cohomology $H^{n+k}(J^{\infty} \mathcal{M}_{M})^{\mathrm{Diff}%
^{e}M}$ of the jet bundle can be related to relative Gelfand-Fuchs cohomology
of of formal vector fields. As the Gelfand-Fuchs cohomology is usually
computed in terms of truncated Weil algebras, let us recall its definition.
Let $\mathfrak{g}$ be a Lie algebra, and denote by $W(\mathfrak{g}%
)=\bigwedge\mathfrak{g}^{*}\otimes S\mathfrak{g}^{*}$ its Weil algebra, where
the elements of $\bigwedge^{r}\mathfrak{g}^{*}$ have degree $r$ and the
elements of $S^{r}\mathfrak{g}^{*}$ have degree $2r$. Let $\{e^{i}\}$ be a
basis of $\mathfrak{g}^{*}$. We denote by $\lambda_{i}=e^{i}\otimes1$ and
$\Lambda_{i}=1\otimes e^{i}$. The Weil algebra $W(\mathfrak{g})$ is a
graduated differential algebra with differential determined by setting
$d\lambda^{i}=\Lambda^{i}-\frac{1}{2}c^{i}_{jk}\lambda^{j}\lambda^{k}$,
$d\Lambda^{i}=-c^{i}_{jk}\Lambda^{j}\lambda^{k}$, and is a classical result
that $W(\mathfrak{g})$ is acyclic. The interior product is defined by setting
$\iota_{e_{i}}\lambda^{j}=\delta^{j}_{i}$, $\iota_{e_{i}}\Lambda^{j}=0$, and
the Lie derivative is defined by $L_{e_{i}}=d\iota_{e_{i}}+\iota_{e_{i}}d$.
The $k$-truncated Weyl algebra $W_{(k)}(\mathfrak{g})$ is the quotient
$W(\mathfrak{g})/J^{k+1}$, where $J^{k}$ is the ideal of $W(\mathfrak{g})$
generated by $S^{k} \mathfrak{g}^{*}$.

If $\mathfrak{h}$ is a Lie subalgebra of $\mathfrak{g}$, we define the
cohomology of $W(\mathfrak{g})$ relative to $h$, $H(W(\mathfrak{g}%
),\mathfrak{h})$ as the cohomology of the $\mathfrak{h}$-basic elements of
$W(\mathfrak{g})$. It is easy to see that $H(W(\mathfrak{g}),\mathfrak{g}%
)\cong I^{G}$ for a connected Lie group $G$.

Let $P\rightarrow M$ be a principal $G$-bundle. A connection form $A$ on $P$
determines a homomorphism of DGA's $w_{A} \colon W(\mathfrak{g})\rightarrow
\Omega(P)$ by mapping $\lambda$ to $A$ and $\Lambda$ to $F_{A}$, the curvature
form of $A$. The induced map in relative cohomology $w_{A} \colon
H(W(\mathfrak{g}),\mathfrak{g})\cong I^{G} \rightarrow\Omega(M)$ coincides
with the Chern-Weil homomorphism. If a Lie group $\mathcal{G}$ acts on $P$ by
automorphisms and $A$ is $\mathcal{G}$-invariant, then the map $w_{A}$ takes
its values on the space of $\mathcal{G}$-invariant forms, $w_{A} \colon
W(\mathfrak{g})\rightarrow\Omega(P)^{\mathcal{G}}$.

As $\mbox{\boldmath$\omega$}_{\mathrm{hor}}$ is a $\mathrm{Diff}(M)$-invariant
connection on the principal $Gl(n)$-bundle $q_{1}^{*}FM\to J^{1}
\mathcal{M}_{M}$, it determines a homomorphism $W(\mathfrak{gl}(n))\rightarrow
\Omega(q_{1}^{*}FM)^{\mathrm{Diff}^{e} M}$. By formula (\ref{Omegahor}) it
factors to a map $W_{(n)}(\mathfrak{gl}(n))\rightarrow\Omega(q_{1}%
^{*}FM)^{\mathrm{Diff}^{e} M}$. By restricting this map to $O^{+}M$ we obtain
a map $W_{(n)}(\mathfrak{gl}(n))\rightarrow\Omega(O^{+}M)^{\mathrm{Diff}^{e}
M}$. Taking into account that $O^{+}M/SO(n)\cong J^{1} \mathcal{M}_{M}$ we
obtain an induced map in relative cohomology $\alpha\colon H(W_{(n)}%
(\mathfrak{gl}(n)),\mathfrak{so}(n))\rightarrow H(J^{1} \mathcal{M}%
_{M})^{\mathrm{Diff}^{e} M}$. Finally, by composing with the projection
$q_{\infty,1}\colon J^{\infty} \mathcal{M}_{M}\to J^{1} \mathcal{M}_{M}$ we
obtain a map $\alpha\colon H(W_{(n)}(\mathfrak{gl}(n)),\mathfrak{so}%
(n))\rightarrow H(J^{\infty} \mathcal{M}_{M})^{\mathrm{Diff}^{e} M}$.

\begin{theorem}
\label{injcoh} The map $\alpha\colon H(W_{(n)}(\mathfrak{gl}(n)),\mathfrak{so}%
(n))\rightarrow H(J^{\infty} \mathcal{M}_{M})^{\mathrm{Diff}^{e} M}$ is injective.
\end{theorem}

The proof of Theorem \ref{injcoh} is given in Section \ref{proof1} and is
based on the ideas explained in \cite{AndPoh} relating the cohomology of
invariant variational bicomplexes to relative cohomology of formal vector fields.

The cohomology $H(W_{(n)}(\mathfrak{gl}(n)),\mathfrak{so}(n))$ is well known
due to its appearance in the cohomology of formal vector fields and
characteristic classes of foliations (e.g. see \cite{godbillon,KT}). Let
$WO_{n}=\bigwedge(U_{1},U_{3},\ldots,U_{2k-1})\otimes S_{n}[C_{1},C_{2}%
,\ldots,C_{n}]$, where $2k-1$ is the greater odd number $\leq n$, $\deg
(U_{i})=2i-1$, $\deg(C_{i})=2i$, and $S_{n}[C_{1},C_{2},\ldots,C_{n}]$ is the
quotient of $S[C_{1},C_{2},\ldots,C_{n}]$ by the ideal $J$ generated by the
elements of degree greater than $2n$. $WO_{n}$ is a differential graded
algebra (DGA) with differential $dU_{i}=C_{i}$, $dC_{i}=0$.  We have (see
\cite{KT}) $H(W_{(n)}(\mathfrak{gl}(n)),\mathfrak{so}(n))\cong H(WO_{n})$ for
$n$ odd, and $H(W_{(n)}(\mathfrak{gl}(n)),\mathfrak{so}(n))\cong
H(WO_{n})[T]/(T^{2}-C_{n})$ for $n$ even. If we set $P_{i}=C_{2i}$, it is easy
to see that for $r\leq2n$ we have $H^{r}(WO_{n})\cong S[P_{1},\ldots
P_{[n/2]}]_{r}$ the space of degree $r$ elements on $S[P_{1},\ldots
,P_{[n/2]}]$. Hence, we conclude that for $r\leq2n$ we have $H^{r}%
(W_{(n)}(\mathfrak{gl}(n)),\mathfrak{so}(n)) \cong I_{r/2}^{SO(n)}$ for $r$
even, and $H^{r}(W_{(n)}(\mathfrak{gl}(n)),\mathfrak{so}(n))=0$ for $r$ odd.

\begin{corollary}
\label{inj1}The map $I_{k}^{SO(n)}\rightarrow H^{2k} (J^{\infty}%
\mathcal{M}_{M})^{\mathrm{Diff}^{e}M}$, $p\mapsto p(\mbox{\boldmath$\Omega$})$
is injective for $k\leq n$. Hence a form $p(\mbox{\boldmath$\Omega$})$ is the
exterior differential of a $\mathrm{Diff}^{e}M$-invariant form on $J^{\infty
}\mathcal{M}_{M}$ if and only if $p=0$.
\end{corollary}

\begin{proof}
In $q_{1}^{\ast}FM$ we have considered two $\mathrm{Diff}M$-invariant
connections, $\mbox{\boldmath$\omega$}_{\mathrm{hor}}$ and
$\mbox{\boldmath$\omega$}$, the second one being a Riemannian connection. If
$p\in I_{r}^{O(n)}$ then we have $p(\mbox{\boldmath$\Omega$})-
p(\mbox{\boldmath$\Omega$} _{\mathrm{hor}})= d(Tp(\mbox{\boldmath$\omega$},
\mbox{\boldmath$\omega$} _{\mathrm{hor}}))$, where
$Tp(\mbox{\boldmath$\omega$}, \mbox{\boldmath$\omega$} _{\mathrm{hor}}%
)\in\Omega^{2r-1} (J^{\infty}\mathcal{M}_{M})^{\mathrm{Diff}M}$ is the
transgression form corresponding to $\mbox{\boldmath$\omega$}_{\mathrm{hor}}$
and $\mbox{\boldmath$\omega$}$ (see \cite{KN}). As $\mbox{\boldmath$\omega$}
_{\mathrm{hor}}$ and $\mbox{\boldmath$\omega$}$ are both $\mathrm{Diff}%
M$-invariant, the form $Tp(\mbox{\boldmath$\omega$},\mbox{\boldmath$\omega$}
_{\mathrm{hor}})$ is also $\mathrm{Diff}M$-invariant. Hence, the forms
$p(\mbox{\boldmath$\Omega$})$ and $p(\mbox{\boldmath$\Omega$}_{\mathrm{hor}})$
determine the same cohomology class on $H^{\bullet}(J^{\infty}\mathcal{M}%
_{M})^{\mathrm{Diff}^{e}M}$. As we know that the class of
$p(\mbox{\boldmath$\Omega$}_{\mathrm{hor}})$ on $H^{\bullet}(J^{\infty
}\mathcal{M}_{M})^{\mathrm{Diff}^{e}M}$ is not zero, we conclude that the
class of $p(\mbox{\boldmath$\Omega$})$ is also not zero. Moreover, for $n$ odd
the class of $\chi(\mbox{\boldmath$\Omega$})$ is also not zero as we have
$[\chi(\mbox{\boldmath$\Omega$})^{2}]= [p_{n/2}(\mbox{\boldmath$\Omega$})]
\neq0$.
\end{proof}

\begin{remark}
\label{contra} We can use Corollary \ref{inj1} to show that, as commented
before, in general the map $\Im\colon H^{n+1}(J^{\infty}\mathcal{M}%
_{M})^{\mathrm{Diff}^{e}M}  \rightarrow H_{\mathrm{loc}}^{1}(\mathfrak{Met}%
M)^{\mathrm{Diff}^{e}M}$ is not an isomorphism.

Let us suppose that $n=4k-1$ for an integer $k$. Let $p\in I^{SO(n)}_{k}$ and
consider the corresponding universal Pontryagin form
$p(\mbox{\boldmath$\Omega$})\in\Omega^{n+1}(J^{\infty}\mathcal{M}%
_{M})^{\mathrm{Diff}^{e}M}$. By Theorem \ref{inj1} the class of
$p(\mbox{\boldmath$\Omega$})$ in $H^{n+1}(J^{\infty}\mathcal{M}_{M}%
)^{\mathrm{Diff}^{e}M}$ is not zero. However, the class of $\Im
[p(\mbox{\boldmath$\Omega$})]$ in $H_{\mathrm{loc}}^{1} (\mathfrak{Met}%
M)^{\mathrm{Diff}^{e}M}$ vanishes.

This can be seen in the following way. Let $\alpha\in\Omega^{n}(J^{\infty
}\mathcal{M}_{M})$ be a form satisfying $p(\mbox{\boldmath$\Omega$})=d\alpha$.
Of course $\alpha$ is not $\mathrm{Diff}^{e}M$-invariant, but it is weakly
$\mathrm{Diff}^{e}M$-invariant, as for every $X\in\mathfrak{X}(M)$ we have
$L_{\mathrm{pr}\bar{X}}\alpha=d(\mathfrak{J}(X)+\iota_{\mathrm{pr}\bar{X}%
}\alpha))$, where we have used that $\iota_{\mathrm{pr}\bar{X}}p
(\mbox{\boldmath$\Omega$})=d(\mathfrak{J}(X))$ for certain $\mathfrak{J}%
(X)\in\Omega^{n-1}(J^{\infty}\mathcal{M}_{M})$. This fact follows from the
existence of equivariant Pontryagin classes (see \emph{\cite{equiconn}}). For
example for $n=3$ and $p=p_{1}$ the first Pontryagin polynomial we can take
(see \emph{\cite[formula (8)]{equiconn}}) $\mathfrak{J}(X)=\frac{1}{4\pi^{2}%
}\mathrm{tr} \left( (\mbox{\boldmath$\nabla$} X)_{A} \circ
\mbox{\boldmath$\Omega$} \right) $, where $(\mbox{\boldmath$\nabla$} X)_{A}$
denotes the skew-symmetric part of $\mbox{\boldmath$\nabla$} X\in\Omega
^{0}(J^{\infty}\mathcal{M}_{M},\mathrm{End}TM)$ and in local coordinates we
have $\mbox{\boldmath$\nabla$} X= (\frac{\partial X^{i}}{\partial x^{j}}
-\mbox{\boldmath$\Gamma$}^{i}_{jk} X^{k})dx^{j}\otimes\frac{\partial}{\partial
x^{i}} $.
\end{remark}

\begin{remark}
\label{RAnd} The map $\alpha$ of Theorem \ref{inj1} is in fact bijective
(\emph{{\cite{Anderson}}}), i.e. all the cohomology classes on $H(J^{\infty}
\mathcal{M}_{M})^{\mathrm{Diff}^{e} M}$ come from $H(W_{(n)}(\mathfrak{gl}%
(n)),\mathfrak{so}(n))$. Using this result, we obtain that $H_{\mathrm{loc}%
}^{k} (\mathfrak{Met}M)^{\mathrm{Diff}^{e}M} \cong H^{n+k}(W_{(n)}%
(\mathfrak{gl}(n)),\mathfrak{so}(n))$ for $k>1$. Also we have $H_{\mathrm{loc}%
}^{0} (\mathfrak{Met}M)^{\mathrm{Diff}^{e}M}=\mathbb{R}$, and for $k=1$ the
preceding remark shows that $H_{\mathrm{loc}}^{1} (\mathfrak{Met}%
M)^{\mathrm{Diff}^{e}M}=0$. However, we confine ourselves to prove
\emph{Theorem \ref{inj1}}. Note that this result is sufficient for the study
of gravitational anomalies done in \cite{anomalies}.
\end{remark}

\subsection{Gelfand-Fuchs cohomology}

Let us recall some basic results about Gelfand-Fuchs cohomology of formal
vector fields. We refer to \cite{Bott,godbillon} for the details. Let
$\mathfrak{a}_{n}= \{X=X^{i}\partial/\partial x^{i}\colon X^{i}\in
\mathbb{R}[[x_{1},\ldots x_{n}]]\}$ be the Lie algebra of formal vector fields
on $\mathbb{R}^{n}$, with Lie bracket
\[
\left[  X^{i}\frac{\partial}{\partial x^{i}},Y^{i}\frac{\partial}{\partial
x^{i}}\right]  =\left(  X^{j}\frac{\partial Y^{i}}{\partial x^{j}} -Y^{j}
\frac{\partial X^{i}}{\partial x^{j}}\right)   \frac{\partial}{\partial x^{i}%
}.
\]
The Lie algebra $\mathfrak{gl}(n,\mathbb{R})$ can be considered as a
subalgebra of $\mathfrak{a}_{n}$ by the map $\mathfrak{gl}(n,\mathbb{R}%
)\rightarrow\mathfrak{a}_{n}$, $a_{j}^{i}\mapsto a_{j}^{i}x^{j}\partial
/\partial x^{i}$. We define $\mathfrak{a}_{n}^{\ast}$ as the space generated
by
\[
\theta^{i}(X)=X^{i}(0),\quad\theta_{j_{1}\cdots j_{k}}^{i}(X)=(-1)^{k}
\frac{\partial^{r}X^{i}} {\partial x^{j_{1}}\cdots\partial x^{j_{r}}}(0),
\]
and we set $R_{j}^{i}=d\theta_{j}^{i}+\theta_{k}^{i}\wedge\theta_{j}^{k}$. It
can be seen that we have $R_{j}^{i}=\theta^{r}\wedge\theta_{jr}^{i}$.

A DGA's homomorphism $W_{(n)}(\mathfrak{gl}(n))\rightarrow\bigwedge
\mathfrak{a}^{*}_{n}$ is defined by mapping $\lambda^{i}_{j}$ to $\theta
^{i}_{j}$ and $\Lambda^{i}_{j}$ to $R^{i}_{j}$. This map induces isomorphisms
in cohomology $H(W_{(n)}(\mathfrak{gl}(n)))\cong H(\mathfrak{a}_{n})$ and
$H(W_{(n)}(\mathfrak{gl}(n)),\mathfrak{so}(n)) \cong H(\mathfrak{a}%
_{n},\mathfrak{so}(n))$.

\subsection{Proof of Theorem \ref{injcoh}}

\label{proof1} Let us consider a coordinate system $(U;x^{i})$. Using this
local chart, for any $x\in M$, $X\in\mathfrak{X}(M)$ we can identify
$j_{x}^{\infty}X$ with an element of $\mathfrak{a}_{n}$. Let us consider a
point $\sigma=j_{x}^{\infty}g\in J^{\infty}\mathcal{M}_{M}$. For simplicity we
take $\sigma=j_{0}^{\infty}(g_{0})$, where $g_{0}=\sum_{i}dx^{i}\otimes
dx^{i}$, and hence we have $y_{ij}(\sigma)=\delta_{ij}$, $y_{ij,J}(\sigma)=0$
for every multiindex $J$. We define a map $\nu_{\sigma}\colon\mathfrak{a}_{n}
\rightarrow T_{\sigma}J^{\infty}\mathcal{M}_{M}$ in the following way. Given
$Y\in\mathfrak{a}_{n}$ let $X\in\mathfrak{X}(M)$ be such that $j_{x}^{\infty
}X=Y$. Then we set $\nu_{\sigma}(Y)=\mathrm{pr}\bar{X}(\sigma)$,  which is
well defined as $\mathrm{pr}\bar{X}(\sigma)$ only depends on the derivatives
of $X$ at $x$. From (\ref{prX}) and (\ref{expXbarra}) it follows that the
kernel of the map $\nu_{\sigma}$ is identified with the Lie subalgebra
$\mathfrak{so}(n)\subset\mathfrak{a}_{n}$.

According to \cite{AndPoh} we define a map
\begin{align*}
\psi_{\sigma}\colon\Omega^{k}(J^{\infty} \mathcal{M}_{M})^{\mathrm{Diff}%
^{+}M}  &  \rightarrow\Omega^{k}(\mathfrak{a}_{n}, \mathfrak{so}(n))\\
\psi_{\sigma}(\alpha)(Y_{1},\ldots,Y_{k})  &  =(-1)^{k} \alpha_{\sigma}%
(\nu_{\sigma}(Y_{1}), \ldots,\nu_{\sigma}(Y_{k})),
\end{align*}
for $\alpha\in\Omega^{n+2}(J^{\infty} \mathcal{M}_{M})^{\mathrm{Diff}^{+}M}$,
and $Y_{1},\ldots,Y_{n+2}\in\mathfrak{a}_{n}$. It is a cochain map and induces
a map in cohomology $\psi_{\sigma}\colon H^{k}(J^{\infty} \mathcal{M}%
_{M})^{\mathrm{Diff}^{+}M} \rightarrow H^{k}(\mathfrak{a}_{n},\mathfrak{so}%
(n))$. We have the following

\begin{proposition}
\label{commut}The following diagram is commutative
\[%
\begin{array}
[c]{c}%
\quad\quad H(W_{(n)}(\mathfrak{gl}(n)),\mathfrak{so}(n))\\%
\begin{array}
[c]{rcl}%
{\scriptstyle \alpha} \swarrow & \quad & \searrow{\scriptstyle \beta}\\
H(J^{1} \mathcal{M}_{M})^{\mathrm{Diff}^{e} M} & \ \overset{\psi
}{\longrightarrow} & H(\mathfrak{a}_{n},\mathfrak{so}(n)).
\end{array}
\end{array}
\]

\end{proposition}

As the map $\beta$ is an isomorphism, we conclude that $\alpha$ is injective
and $\psi$ is surjective, proving Theorem \ref{injcoh}.

\begin{proof}
As vector spaces we have $\mathfrak{gl}(n)\cong\mathfrak{so}(n)\oplus
\mathrm{sym}(n)$, where $\mathrm{sym}(n)$ denotes the space of symmetric
matrices of order $n$. Hence we have $W(\mathfrak{gl}(n))\cong\bigwedge
\mathfrak{so}(n)^{*}\otimes\bigwedge\mathrm{sym}(n)^{*}\otimes S\mathfrak{gl}%
(n)^{*}$. By definition, the space of $\mathfrak{so}(n)$-horizontal elements
of $W(\mathfrak{gl}(n))$ is $\bigwedge\mathrm{sym}(n)^{*}\otimes
S\mathfrak{gl}(n)^{*}$. If $\lambda^{S}$ denotes the symmetric part of
$\lambda$, the map $\beta$ maps $q(\lambda^{S},\Lambda)\in W(\mathfrak{gl}%
(n))_{\mathfrak{so}(n)\mathrm{-basic}}$ to $q(\theta^{S},R)$

We denote by $\mbox{\boldmath$\omega$}_{\mathrm{hor}}^{A} \in\Omega^{1}%
(q_{1}^{*}FM,\mathfrak{so}(n))$ and $\mbox{\boldmath$\omega$}_{\mathrm{hor}%
}^{S} \in\Omega^{1}(q_{1}^{*}FM,\mathrm{sym}(n))$ the skew-symmetric and
symmetric parts of the connection $\mbox{\boldmath$\omega$}_{\mathrm{hor}}
\in\Omega^{1}(q_{1}^{*}FM,\mathfrak{gl}(n))$ respectively. As we have
$\mbox{\boldmath$\omega$}_{\mathrm{hor}} =\mbox{\boldmath$\omega$}-\frac{1}%
{2}\mbox{\boldmath$\vartheta$}$ and $\mbox{\boldmath$\omega$}\in\Omega
^{1}(q_{1}^{*}FM,\mathfrak{so}(n))$, $\mbox{\boldmath$\vartheta$}\in\Omega
^{1}(q_{1}^{*}FM,\mathrm{sym}(n))$, we clearly have
$\mbox{\boldmath$\omega$}_{\mathrm{hor}}^{A} =\mbox{\boldmath$\omega$}$,
$\mbox{\boldmath$\omega$}_{\mathrm{hor}}^{S} =-\frac{1}{2}%
\mbox{\boldmath$\vartheta$}$.

By the definition of $\alpha$, $q(\lambda^{S},\Lambda)\in W(\mathfrak{gl}%
(n))_{\mathfrak{so}(n)\mathrm{-basic}}$ is mapped by $\alpha$ to
$q(\mbox{\boldmath$\omega$}_{\mathrm{hor}}^{S},
\mbox{\boldmath$\Omega$}_{\mathrm{hor}})$. From formulas (\ref{expXbarra}),
(\ref{Omegahor}) and (\ref{vartheta}) it follows the following

\begin{lemma}
\label{OmegaR} If $X=X^{i}\partial/\partial x^{i}$, $Y=Y^{i}\partial/\partial
x^{i}$ are the local expressions of two vector fields on $M$, then we have
\[
((\mbox{\boldmath$\Omega$}_{\mathrm{hor}})_{j_{x}^{\infty}g} (\mathrm{pr}%
\bar{X},\mathrm{pr}\bar{Y}))_{j}^{i}= X^{r}(x)\frac{\partial^{2}Y^{i}}
{\partial x^{j}\partial x^{r}}(x)-Y^{r}(x)\frac{\partial^{2}X^{i}} {\partial
x^{j}\partial x^{r}}(x) =R^{i}_{j}(X,Y).
\]

\[
((\mbox{\boldmath$\omega$}_{\mathrm{hor}}^{S})_{j_{x}^{\infty}g}
(\mathrm{pr}\bar{X}))_{j}^{i}=\! -\frac{1}{2}%
(\mbox{\boldmath$\vartheta$}_{j_{x}^{\infty}g} (\mathrm{pr}\bar{X}))_{j}%
^{i}=\! -\frac{1}{2}\!\left( \!\frac{\partial X^{j}}{\partial x^{i}%
}(x)\!+\!\frac{\partial X^{i}}{\partial x^{j}}(x)\!\right) \!\!=\!-(\theta
^{S})^{i}_{j}(X).
\]

\end{lemma}

We conclude from the preceding Lemma that $\psi_{\sigma}%
(q(\mbox{\boldmath$\omega$}_{\mathrm{hor}}^{S},
\mbox{\boldmath$\Omega$}_{\mathrm{hor}}))=q(\theta^{S},R)$, and this proves
Proposition \ref{commut}.
\end{proof}

\section{Connections and metrics}

\subsection{The universal characteristic forms on the bundle of connections}

Let $\pi\colon P\rightarrow M$ a principal $G$-bundle over a compact
$n$-manifold $M$. We denote by $\mathcal{A}_{P}$ the space of principal
connections on $P$. In order to apply our general constructions about local
cohomology to the case of connections on principal bundles we consider a
bundle (the bundle of connections) $p\colon C(P)\rightarrow M$ whose global
sections correspond to principal connections on $P$, i.e, we have
$\mathcal{A}_{P}\cong\Gamma(M,C(P))$. Let us recall the definition of this
bundle (see \cite{geoconn,Gar2,MS} for details). Let $\bar{p}\colon
J^{1}P\rightarrow P$ be the first jet bundle of $P$. The action of $G$ on $P$
lifts to an action on $J^{1}P$. We denote by $p\colon C(P)=J^{1}P/G\rightarrow
M=P/G$ the quotient bundle, called the bundle of connections of $P$. The
projection ${\bar{\pi}\colon} J^{1}P\rightarrow C(P)$ is a principal
$G$-bundle, isomorphic to the pull-back bundle $p^{\ast}P\rightarrow C(P)$,
that we denote by ${\bar{\pi}\colon}\mathbb{P}\rightarrow C(P)$. We have the
following commutative diagram
\[%
\begin{array}
[c]{ccc}%
\mathbb{P} & \overset{\bar{p}}{\longrightarrow} & P\\
{\scriptstyle\bar{\pi}}\downarrow\text{ \ } &  & \text{ }\downarrow
{\scriptstyle\pi}\\
C(P) & \overset{p}{\longrightarrow} & M
\end{array}
\]
The map $\overline{p}$ is $G$-equivariant, i.e., is a principal $G$-bundle morphism.

The group $\mathrm{Aut}P$ of principal $G$-bundle automorphisms is denoted by
$\mathrm{Aut}P$. If $\phi\in\mathrm{Aut}P$, we denote by $\underline{\phi}%
\in\mathrm{Diff}M$ its projection to $M$. We denote by $\mathrm{Aut}^{+}P$ the
subgroup of elements $\phi\in\mathrm{Aut}P$ such that $\underline{\phi}%
\in\mathrm{Diff}^{+}M$. The kernel of the projection $\mathrm{Aut}%
P\rightarrow\mathrm{Diff}M$ is the gauge group of $P$, denoted by
$\mathrm{Gau}P$.

The Lie algebra of $\mathrm{Aut}P$ can be identified with the space
$\mathrm{aut}P\subset\mathfrak{X}(P)$ of $G$-invariant vector fields on $P$.
The subspace of $G$-invariant vertical vector fields is denoted by
$\mathrm{gau}P$ and can be considered as the Lie algebra of $\mathrm{Gau}P$.
We have an exact sequence of Lie algebras $0\rightarrow\mathrm{gau}P
\rightarrow\mathrm{aut}P \rightarrow\mathfrak{X}(M)\rightarrow0$.

The action of $\mathrm{Aut}P$ on $P$ induces actions on $J^{1}P$ and $C(P)$,
and the maps ${\bar{\pi}}$ and $\bar{p}$ are $\mathrm{Aut}P$-invariant. At the
infinitesimal level, if $X\in\mathrm{aut}P$, we denote by $\underline{X}%
\in\mathfrak{X}(M)$ its projection to $M$, and by $X_{\mathbb{P}}%
\in\mathfrak{X}(\mathbb{P})$, $X_{C(P)}\in\mathfrak{X}(C(P))$ its lifts to
$\mathbb{P}=J^{1}P$ and $C(P)$ respectively.

Let $(U,x^{i})$ be a local coordinate system on $M$, $(B^{\alpha})$ a basis
for $\mathfrak{g}$. If $\tilde{B}_{\alpha}$ denotes the $G$-invariant vector
field $(B_{\alpha})_{P}$, then for every $X\in\mathrm{aut}P$ we have
$X=f^{i}\partial/\partial x^{i}+g^{\alpha} \tilde{B}_{\alpha}$, with
$f^{i},g^{\alpha}\in C^{\infty}(U)$. Let $(p^{-1}U,x^{i},A_{i}^{\alpha})$ be
the induced coordinate system on $C(P)$ (see \cite[Section 3.2]{geoconn}). If
$X\in\mathrm{aut}P$ is given in local coordinates by $X= f^{i}\partial
/\partial x^{i}+g^{\alpha}\tilde{B}_{\alpha}$ then we have%

\begin{equation}
X_{C(P)}=f^{j}\frac{\partial}{\partial x^{j}}- \left(  \frac{\partial f^{i}%
}{\partial x^{j}}A_{i}^{\alpha}+ \frac{\partial g^{\alpha}}{\partial x^{j}}-
c_{\beta\gamma}^{\alpha}g^{\beta}A_{j}^{\gamma}\right)  \frac{\partial
}{\partial A_{j}^{\alpha}},\label{XC}%
\end{equation}
where $c_{\beta\gamma}^{\alpha}$ are the structure constants of $\mathfrak{g}$.

The principal $G$-bundle ${\bar{\pi}\colon}\mathbb{P\rightarrow} C(P)$ is
endowed with a canonical $\mathrm{Aut}P$-invariant connection $\mathbb{A}%
\in\Omega^{1}(\mathbb{P},\mathfrak{g})$. This connection can be identified to
the contact form on $J^{1}P$. Alternatively, it can be defined by setting
$\mathbb{A}_{(\sigma_{A}(x),u)}(X)=A_{u}(\bar{p}_{\ast}X)$, for every
connection $A$ on $P$, $x\in M$, $u\in\pi^{-1}(x)$, $X\in T_{(\sigma
_{A}(x),u)}\mathbb{P}$, and where $\sigma_{A}\colon M\rightarrow C(P)$ is the
section of $C(P)$ corresponding to  $A$. Let $\mathbb{F}$ be the curvature of
$\mathbb{A}$. In local coordinates we have (see \cite{geoconn})
\begin{equation}
\label{F}\mathbb{F}= \left(  dA_{j}^{\alpha}\wedge dx^{j}+ c_{\beta\gamma
}^{\alpha}A_{j}^{\beta}A_{k}^{\gamma}dx^{j} \wedge dx^{k}\right)
\otimes\tilde{B}_{\alpha}.
\end{equation}
If $f\in I_{k}^{G}$ is a Weil polynomial of degree $k$ for $G$, we define the
universal characteristic form associated to $f$ as the $2$-form on $C(P)$
defined by $f(\mathbb{F})= f(\mathbb{F},\ldots,\mathbb{F})\in\Omega
^{2k}(C(P))$.

\subsection{Local cohomology of $\mathfrak{Met}M\times\mathcal{A}_{P}$}

Now we consider the product bundle $\mathcal{M}_{M}\times_{M}C(P)$, whose
space of sections is the product $\mathfrak{Met}M\times\mathcal{A}_{P}$. The
group $\mathrm{Aut}(P)$ acts on $C(P)$ as explained above, and acts on
$\mathcal{M}_{M}$ through its projection $\mathrm{Aut}(P)\to\mathrm{Diff}(M)$.

The connection $\mbox{\boldmath$\omega$}_{\mathrm{hor}}$ determines a torsion
free $\mathrm{Aut}^{+}P$-invariant connection on the space of total vector
fields on $J^{\infty}(\mathcal{M}_{M}\times_{M}C(P))$, and by the results
explained in Section \ref{VB-GF} we have for $k>1$ the isomorphism
\begin{equation}
\label{isocoho2}H_{\mathrm{loc}}^{k} (\mathfrak{Met}M\times\mathcal{A}%
_{P})^{\mathrm{Aut}^{+}P} \cong H^{n+k}(J^{\infty}(\mathcal{M}_{M} \times
_{M}C(P)))^{\mathrm{Aut}^{+}P}.
\end{equation}
As the connection $\mbox{\boldmath$\omega$}_{\mathrm{hor}}\times\mathbb{A}$ is
$\mathrm{Aut}(P)$-invariant, it determines a homomorphism $W(\mathfrak{gl}%
(n)\times\mathfrak{g}) \rightarrow\Omega(q_{1}^{*}FM \times P)^{\mathrm{Aut}%
^{e} P}$. By formulas (\ref{Omegahor}) and (\ref{F}) it factors to a map
$W_{(n)}(\mathfrak{gl}(n)\times\mathfrak{g}) \rightarrow\Omega(q_{1}%
^{*}FM)^{\mathrm{Aut}^{e} M}$. By composing with the inclusion of $O^{+}M$ on
$q_{1}^{*}FM$ we obtain a map $W_{(n)}(\mathfrak{gl}(n)\times\mathfrak{g})
\rightarrow\Omega(O^{+}M \times P)^{\mathrm{Aut}^{e} P}$. Taking into account
that $(O^{+}M\times P)/(SO(n)\times G) \cong J^{1} \mathcal{M}_{M} \times C(P)
$ we obtain a map $\alpha\colon H(W_{(n)}(\mathfrak{gl}(n)\times\mathfrak{g}),
\mathfrak{so}(n)\times\mathfrak{g})\rightarrow H(J^{\infty} (\mathcal{M}%
_{M}\times_{M}C(P)))^{\mathrm{Aut}^{e} P}$.

\begin{theorem}
\label{inj2} The map
\begin{equation}
\alpha\colon H(W_{(n)}(\mathfrak{gl}(n) \times\mathfrak{g}), \mathfrak{so}(n)
\times\mathfrak{g})\rightarrow H(J^{\infty} (\mathcal{M}_{M}\times
_{M}C(P)))^{\mathrm{Aut}^{e} P}%
\end{equation}
is injective.
\end{theorem}

The proof of this Theorem is similar to that of Theorem \ref{inj1} and is
given in Section \ref{proof2}.

The cohomology $H(W_{(n)}(\mathfrak{gl}(n) \times\mathfrak{g}), \mathfrak{so}%
(n) \times\mathfrak{g})$ is computed in \cite{Hamasuki}. In particular for
$k\leq n$ we have $H^{2k}(W_{(n)}(\mathfrak{gl}(n) \times\mathfrak{g}),
\mathfrak{so}(n) \times\mathfrak{g})\cong\bigoplus_{r+s=k}I_{r}^{SO(n)}
{\textstyle\bigotimes} I_{s}^{G}.$ Hence we have the following

\begin{corollary}
\label{cor3} The map $\bigoplus_{r+s=k}I_{r}^{SO(n)} {\textstyle\bigotimes}
I_{s}^{G} \! \rightarrow\! H^{2k}(J^{\infty}(\mathcal{M}_{M}\times
_{M}C(P)))^{\mathrm{Aut}^{+}P}$, $p\otimes f \mapsto\lbrack p(
\mbox{\boldmath$\Omega$}) \wedge f(\mathbb{F})]$ is injective for $k\leq n$.
\end{corollary}

We have also the following

\begin{corollary}
The map $I_{k}^{G} \rightarrow H^{2k}(J^{\infty}(C(P)))^{\mathrm{Aut}^{+}P}$,
$f\mapsto\lbrack f(\mathbb{F})]$ is injective for $k\leq n$.
\end{corollary}

\subsection{Cohomology of formal $G$-invariant vector fields}

Let $G$ be a Lie group with Lie algebra $\mathfrak{g}$. We consider a basis
${B^{\alpha}}$ of $\mathfrak{g}$, and we denote by $c^{\alpha}_{\beta\gamma}$
the structure constants of $\mathfrak{g}$ in this basis.

Let $\mathfrak{a}_{n,\mathfrak{g}}= \{f^{i}\partial/\partial x^{i}+ g^{\alpha
}B_{\alpha}\colon f^{i}, g^{\alpha}\in\mathbb{R}[[x_{1},\ldots x_{n}]] \}$ be
the Lie algebra of formal $G$-invariant vector fields on $\mathbb{R}^{n}\times
G$, with Lie bracket given by%

\begin{align*}
\left[  f^{i}\frac{\partial}{\partial x^{i}},k^{i} \frac{\partial}{\partial
x^{i}}\right]   &  =\left(  f^{j}\frac{\partial k^{i}} {\partial x^{j}}-k^{j}
\frac{\partial f^{i}}{\partial x^{j}}\right)  \frac{\partial}{\partial x^{i}%
},\\
\left[  f^{i}\frac{\partial}{\partial x^{i}}, g^{\alpha}B_{\alpha}\right]   &
=f^{i}\frac{\partial g^{\alpha}}{\partial x^{i}} B_{\alpha},\quad\quad\left[
g^{\alpha}B_{\alpha},h^{\alpha}B_{\alpha}\right]  =c_{\beta\gamma}^{\alpha
}g^{\beta}h^{\gamma}B_{\alpha},
\end{align*}

We define $\mathfrak{a}_{n,\mathfrak{g}}^{\ast}$ as the space generated by
\begin{align*}
\theta^{i}(X)  &  =f^{i}(0),\quad\theta_{j_{1}\cdots j_{k}}^{i}(X)=(-1)^{k}
\frac{\partial^{k} f^{i}}{\partial x^{j_{1}}\cdots\partial x^{j_{k}}}(0),\\
\sigma^{\alpha}(X)  &  =g^{\alpha}(0),\quad\sigma_{j_{1}\cdots j_{k}}^{\alpha
}(X)=(-1)^{k}\frac{\partial^{k} g^{\alpha}}{\partial x^{j_{1}}\cdots\partial
x^{j_{k}}}(0),
\end{align*}
and we set $R_{j}^{i}=d\theta_{j}^{i}+\theta_{k}^{i}\wedge\theta_{j}^{k}$ and
$S^{\alpha}=d\sigma^{\alpha}+\tfrac{1}{2}c_{\beta\gamma}^{\alpha}\sigma
^{\beta}\sigma^{\gamma}$. We have (see \cite{Hamasuki})
\begin{align}
R_{j}^{i}  &  =\theta^{k}\wedge\theta_{jk}^{i},\quad\quad dR_{j}^{i}
+\theta_{k}^{i}\wedge R_{j}^{k}-R_{k}^{i}\wedge\theta_{j}^{k}=0,\label{R}\\
S^{\alpha}  &  =\theta^{i}\wedge\sigma_{i}^{\alpha},\quad\quad dS^{\alpha
}+c_{\beta\gamma}^{\alpha}S^{\beta}\sigma^{\gamma}=0.\label{S}%
\end{align}

The Lie algebra $\mathfrak{gl}(n)$ is considered as a Lie subalgebra of
$\mathfrak{a}_{n,\mathfrak{g}}$ through the map $\mathfrak{gl}(n)\rightarrow
\mathfrak{a}_{n,\mathfrak{g}}$, $A_{j}^{i}\mapsto A_{j}^{i}x^{j}%
\partial/\partial x^{i}$, and similarly $\mathfrak{g}$ is considered as a Lie
subalgebra of $\mathfrak{a}_{n,\mathfrak{g}}$ with the obvious map.

The Lie algebra cohomology of $\mathfrak{a}_{n,\mathfrak{g}}$ relative to a
Lie subalgebra $\mathfrak{h}$ is the cohomology of the subcomplex of
$\mathfrak{h}$-basic elements in $\bigwedge\mathfrak{a}_{n,\mathfrak{g}}$, and
is denoted by $H(\mathfrak{a}_{n,\mathfrak{g}},\mathfrak{h})$. As usual, the
cohomology of $\mathfrak{a}_{n,\mathfrak{g}}$ can be computed in terms of the
cohomology of truncated Weil algebras. We consider the Weil algebra
$W(\mathfrak{gl}(n)\times\mathfrak{g})$ of the Lie algebra $\mathfrak{gl}%
(n)\times\mathfrak{g}$. By formulae (\ref{R}) and (\ref{S}) we have a map
$\beta\colon W_{(n)}(\mathfrak{gl}(n,\mathbb{R})\times\mathfrak{g})
\rightarrow\bigwedge\mathfrak{a}_{n,\mathfrak{g}}$ defined by setting
$\beta(\lambda^{i}_{j})=\theta^{i}_{j}$, $\beta(\Lambda^{i}_{j})=R^{i}_{j}$,
$\beta(\lambda^{\alpha})=\sigma^{\alpha}$ and $\beta(\Lambda^{\alpha
})=S^{\alpha}$. In \cite{Hamasuki} it is proved that the induced map on
relative cohomology $\beta\colon H^{k}(\mathfrak{a}_{n,\mathfrak{g}%
},\mathfrak{so}(n)\times\mathfrak{g}) \to H^{k}(W_{(n)}(\mathfrak{gl}%
(n,\mathbb{R})\times\mathfrak{g}), \mathfrak{so}(n)\times\mathfrak{g}) $ is an isomorphism.

\subsection{Proof of Theorem \ref{inj2}\label{proof2}}

Let us fix a local trivialization $(\pi^{-1}(U),x^{i},u^{\alpha})$ of $P$ such
that $x^{i}(x)=0$ and a point $\sigma=(j_{x}^{\infty}g,j_{x}^{\infty}A)\in
J^{\infty}(\mathcal{M}_{M}\times_{M}C(P))$. Again for simplicity we assume
that $y_{ij}(j_{x}^{\infty}g)=\delta_{ij}$, $y_{ij,J}(j_{x}^{\infty}g)=0$, and
$A_{j,J}^{\alpha}(j_{x}^{\infty}A)=0$, for every multiindex $J$. Using this
trivialization, for any $X\in\mathrm{aut}P$ we can identify $j_{x}^{\infty}X$
with an element of $\mathfrak{a}_{n,\mathfrak{g}}$. We define a map
$\nu_{\sigma}\colon\mathfrak{a}_{n,\mathfrak{g}} \rightarrow T_{\sigma
}J^{\infty} (\mathcal{M}_{M}\times_{M}C(P))$ in the following way. Given
$Y\in\mathfrak{a}_{n,\mathfrak{g}}$ let $X\in\mathrm{aut}P$ be such that
$j_{x}^{\infty}X=Y$. Then we set $\nu_{\sigma}(Y)= \mathrm{pr}(\bar
{X},X_{C(P)})(\sigma)$, which is well defined as $\mathrm{pr}(\bar{X}%
,X_{C(P)})(\sigma)$ only depends on the derivatives of $X$ at $x$. From
(\ref{prX}), (\ref{expXbarra}) and (\ref{XC}) if follows that the kernel of
the map $\nu_{\sigma}$ is identified with the Lie subalgebra $\mathfrak{so}%
(n)\times\mathfrak{g}\subset\mathfrak{a}_{n,\mathfrak{g}}$.

According to \cite{AndPoh} we define a map
\[
\psi_{\sigma}\colon H^{k}(J^{\infty} (\mathcal{M}_{M}\times_{M}%
C(P)))^{\mathrm{Aut}^{+}P} \rightarrow H^{k}(\mathfrak{a}_{n,\mathfrak{g}}
,\mathfrak{so}(n)\times\mathfrak{g})
\]
by setting $\psi_{\sigma}(\alpha)(Y_{1},\ldots,Y_{k})=(-1)^{k}\alpha_{\sigma}
(\nu_{\sigma}(Y_{1}),\ldots,\nu_{\sigma}(Y_{k}))$\ for $Y_{1},\ldots,Y_{k}%
\in\mathfrak{a}_{n,\mathfrak{g}}$, and $\alpha\in\Omega^{k}(J^{\infty}
(\mathcal{M}_{M}\times_{M}C(P)))^{\mathrm{Aut}^{+}P}$.

\begin{proposition}
\label{commut2} The following diagram is commutative
\[%
\begin{array}
[c]{c}%
\quad\quad H(W_{(n)}(\mathfrak{gl}(n) \times\mathfrak{g}), \mathfrak{so}(n)
\times\mathfrak{g})\\%
\begin{array}
[c]{rcl}%
{\scriptstyle \alpha} \swarrow & \quad & \searrow{\scriptstyle \beta}\\
H(J^{\infty} (\mathcal{M}_{M} \times C(P)))^{\mathrm{Aut}^{e} P} &
\ \overset{\psi}{\longrightarrow} & H(\mathfrak{a}_{n,\mathfrak{g}%
},\mathfrak{so}(n)\times\mathfrak{g})
\end{array}
\end{array}
\]

\end{proposition}

As the map $\beta$ is an isomorphism, we conclude that $\alpha$ is injective,
proving Theorem \ref{inj2}.

The proof of Proposition \ref{commut2} is the same than that of Proposition
\ref{commut}, using Lemma \ref{OmegaR} and the following lemma, that shows
that $\psi_{\sigma}([p(\mbox{\boldmath$\Omega$}_{\mathrm{hor}}) \wedge
f(\mathbb{F})])=[p(R)\wedge f(S)]$.

\begin{lemma}
\label{FS} If $X=f_{1}^{i}\partial/\partial x^{i}+g_{1}^{\alpha}\tilde
{B}_{\alpha}$, $Y=f_{2}^{i}\partial/\partial x^{i}+g_{2}^{\alpha}\tilde
{B}_{\alpha}$, is the local expression of $X,Y\in\mathrm{aut}P$, we have
\[
\mathbb{F}(X_{C(P)},Y_{C(P)})=\left(  f_{1}^{i}\partial g_{2}^{\alpha
}/\partial x^{i}-f_{2}^{i}\partial g_{1}^{\alpha}/\partial x^{i}\right)
\otimes\tilde{B}_{\alpha}.
\]

\end{lemma}

\begin{acknowledgement}
I would like to thank I. Anderson for letting me know some of his unpublished
results on the cohomology of invariant variational bicomplexes, and to P.
Martinez Gadea for calling my attention to reference \cite{Hamasuki}. This
work is supported by Ministerio de Educaci\'{o}n y Ciencia of Spain, under
grant \emph{\#MTM2005--00173}.
\end{acknowledgement}

\newpage

\part{Local Anomalies and Local Equivariant Cohomology}
\begin{abstract}
The locality conditions for the vanishing of local anomalies in
field theory are shown to admit a geometrical interpretation in
terms of local equivariant cohomology. This interpretation allows
us to solve the problem proposed by Singer in \cite{singer}, and
consisting in defining an adequate notion of local cohomology to
deal with the problem of locality in the geometrical approaches to
the study of local anomalies based on the Atiyah-Singer index
theorem. Moreover, using the relation between local cohomology and
the cohomology of jet bundles studied in \cite{localVB} we obtain
necessary and sufficient conditions for the cancellation of local
gravitational and mixed anomalies.

\end{abstract}
\noindent\emph{Key words and phrases:} local equivariant cohomology, local
anomalies, equivariant characteristic classes, BRST cohomology.

\smallskip

\noindent\emph{Mathematics Subject Classification 2000:} Primary
81T50; Secondary 55N91, 57R20, 58D17, 58A20, 70S15.
\setcounter{section}{0} \setcounter{theorem}{0}
\section{Introduction\label{introduction}}

An anomaly appears in a theory when a classical symmetry is broken at the
quantum level. One fundamental concept in the study of local anomalies is
locality. In order to cancel the anomaly, only local terms are allowed,
``local'' meaning terms obtained integrating forms depending on the fields and
its derivatives. In the algebraic approaches to local anomalies (local BRST
cohomology, descent equations) only local terms are considered. However, in
the geometric and topological approaches based on the Atiyah-Singer index
theorem it is not clear how to deal with the problem of locality. The aim of
the present paper is to solve the old problem, suggested by Singer in
\cite{singer}, and consisting in determining an adequate notion of ``local
cohomology'' which allows to deal with the problem of locality in that
geometric approaches.

Let us briefly recall some basic ideas about the problem of locality in the
study of local anomalies (e.g. see \cite{ber}). In this paper we consider only
local anomalies, and hence we can assume that we are dealing with a connected
group $\mathcal{G}$, with Lie algebra $\mathfrak{G}$. We consider an action of
$\mathcal{G}$ on a bundle $E\rightarrow M$ over a compact $n$-manifold $M$.
Let $\{D_{s}:s\in\Gamma(E)\}$ be a $\mathcal{G}$-equivariant family of
elliptic operators acting on fermionic fields $\psi\in\Gamma(V)$ and
parametrized by $\Gamma(E)$. Then the lagrangian density $\mathcal{L}%
(\psi,s)=\bar{\psi}iD_{s}\psi$ is $\mathcal{G}$-invariant, and hence the
classical action $S_{\mathcal{L}} (\psi,s)=\int_{M}\mathcal{L}(\psi,s)$, is a
$\mathcal{G}$-invariant function on $\Gamma(V)\times\Gamma(E)$. However, at
the quantum level, the corresponding effective action $W(s)$, defined in terms
of the fermionic path integral by $\exp(-W(s))=\int\mathcal{D}\psi
\mathcal{D}\bar{\psi}\exp\left(  -\int_{M}\bar{\psi}iD_{s}\psi\right) $ could
fail to be $\mathcal{G}$-invariant if the fermionic measure $\mathcal{D}%
\psi\mathcal{D}\bar{\psi}$ is not $\mathcal{G}$-invariant. To measure this
lack of invariance we define $\mathcal{A}\in\Omega^{1}(\mathfrak{G},\Omega
^{0}(\Gamma(E)))$ by $\mathcal{A}=\delta W$, i.e.$\ \mathcal{A}(X)(s)=L_{X}%
W(s)$ for $X\in\mathfrak{G}$, $s\in\Gamma(E)$. Although $W$ is clearly a
non-local functional, $\mathcal{A}$ is local in $X$ and $s$, i.e. we have
$\mathcal{A}\in\Omega_{\mathrm{loc}}^{1} (\mathfrak{G},\Omega_{\mathrm{loc}%
}^{0}(\Gamma(E)))$. It is clear that $\mathcal{A}$ satisfies the condition
$\delta\mathcal{A}=0$ (the Wess-Zumino consistency condition). Moreover, if
$\mathcal{A}=\delta\Lambda$ for a \emph{local} functional $\Lambda=\int
_{M}\lambda\in\Omega_{\mathrm{loc}}^{0}(\Gamma(E))$  then we can define a new
lagrangian density $\mathcal{\hat{L}}=\mathcal{L}+\lambda$, such that the new
effective action $\hat{W}$ is $\mathcal{G}$-invariant, and in that case the
anomaly cancels. If $\mathcal{A}\neq\delta\Lambda$ for every $\Lambda\in
\Omega_{\mathrm{loc}}^{0}(\Gamma(E))$ then we say that there exists an anomaly
in the theory. Hence the anomaly is measured by the cohomology class of
$\mathcal{A}$ in the BRST cohomology $H_{\mathrm{loc}}^{1} (\mathfrak{G}%
,\Omega_{\mathrm{loc}}^{0}(\Gamma(E)))$. In this way the problem of anomaly
cancellation can be reduced to the pure algebraic computation of the BRST
cohomology (e.g. see \cite{BBH,bonora,bonora2,DVHTV,DVTVa,DVTVb,MSZ,schmid}).

Local anomalies also admit a nice geometrical interpretation in terms of the
Atiyah-Singer index theorem for families of elliptic operators (see
\cite{ASZ,AGG,AS,freed,singer}). The first Chern class $c_{1}\left(
\mathrm{det\,Ind}D/\mathcal{G}\right) $ of the (quotient) determinant line
bundle $\mathrm{det\,Ind}D/\mathcal{G}\rightarrow\Gamma(E)/\mathcal{G}$
represents an obstruction for anomaly cancellation. The Atiyah-Singer index
theorem for families provides an explicit expression for $c_{1}\left(
\mathrm{det\,Ind}D/\mathcal{G}\right) $ and more precisely, of the curvature
$\Omega^{\mathrm{det\,Ind}D/\mathcal{G}}$ of its natural connection. Now the
problem of locality appears again. The condition $c_{1}\left(
\mathrm{det\,Ind}D/\mathcal{G}\right) =0$ is a necessary but not a sufficient
condition for local anomaly cancellation. For example (see \cite{ASZ}), for
$M=S^{6}$ although $c_{1}\left( \mathrm{det\,Ind}{\not \! \partial
}/\mathrm{Diff}^{0}M\right) =0$, the local gravitational anomaly does not
cancel. Moreover we recall (see \cite{AS,blau,MR}) that the BRST and index
theory approaches are related by means of the transgression map (see Section
\ref{trans}) $t\colon H^{2}(\Gamma(E)/\mathcal{G}) \rightarrow H^{1}%
(\mathfrak{G},\Omega^{0}(\Gamma(E)))$ \emph{i.e.} $[\mathcal{A}]=t(c_{1}%
\left(  \mathrm{det\,Ind}D/\mathcal{G}\right)  )$. As the transgression map
$t$ is injective, the condition $c_{1}\left(  \mathrm{det\,Ind}D/\mathcal{G}%
\right)  \!=\!0$ on $H^{2}(\Gamma(E)/\mathcal{G})$ is equivalent to
$[\mathcal{A}]\!=\!0$ on $H^{1}(\mathfrak{G},\Omega^{0}(\Gamma(E)))$. However,
the condition for local anomaly cancellation is $[\mathcal{A}]=0$ on the BRST
cohomology $H_{\mathrm{loc}}^{1}(\mathfrak{G},\Omega_{\mathrm{loc}}^{0}%
(\Gamma(E)))$. Hence, in order to cancel the local anomaly, $\Omega
^{\mathrm{det\,Ind}D/\mathcal{G}}$ should be the exterior differential of a
``local'' form on $\Gamma(E)/\mathcal{G}$, and the local anomaly cancellation
should be expressed in terms of an adequate notion of ``local cohomology of
$\Gamma(E)/\mathcal{G}$'', $H_{\mathrm{loc}}^{k}(\Gamma(E)/\mathcal{G})$. Note
however that it is by no means clear how to define $H_{\mathrm{loc}}%
^{k}(\Gamma(E)/\mathcal{G})$, as the expression of $\Omega^{\mathrm{det\,Ind}%
D/\mathcal{G}}$ itself contains non-local terms (Green operators). The problem
of defining this notion of ``local cohomology'' was proposed in \cite{singer}.
In \cite{ASZ} a paper studying the preceding problem is announced to be in
preparation, but to the best of our knowledge, this paper has not been published.

Let us explain how local $\mathcal{G}$-equivariant cohomology solves that
problem. The $\mathcal{G}$-equivariant cohomology of $\Gamma(E)$ and the
cohomology of $\Gamma(E)/\mathcal{G}$ are related by the generalized
Chern-Weil homomorphism $\mathrm{ChW}\colon H_{\mathcal{G}}^{2}(\Gamma(E))
\rightarrow H^{2}(\Gamma(E)/\mathcal{G})$. We define another injective
transgression map $\tau\colon H_{\mathcal{G}}^{2}(\Gamma(E))\!\rightarrow\!
H^{1}(\mathfrak{G},\Omega^{0}(\Gamma(E)))$ in such a way that $t\circ
\mathrm{ChW}=\tau$ (see Section \ref{trans}).

Now, to deal with the problem of locality, we define the local $\mathcal{G}%
$-equivariant cohomology $H_{\mathcal{G},\mathrm{loc}}^{k}(\Gamma(E))$ in a
natural way, and we prove that the restriction of $\tau$ to $H_{\mathcal{G}%
,\mathrm{loc}}^{k}(\Gamma(E))$ takes values on $H_{\mathrm{loc}}^{1}
(\mathfrak{G},\Omega_{\mathrm{loc}}^{0}(\Gamma(E)))$. We set $H_{\mathrm{loc}%
}^{2}(\Gamma(E)/\mathcal{G})= \mathrm{ChW}(H_{\mathcal{G},\mathrm{loc}}%
^{k}(\Gamma(E)))$ and we have the following commutative diagram%

\[%
\begin{array}
[c]{c}%
\begin{array}
[c]{ccc}%
H_{\mathcal{G},\mathrm{loc}}^{2}(\Gamma(E)) & \overset{\mathrm{ChW}%
}{\longrightarrow} & H_{\mathrm{loc}}^{2}(\Gamma(E)/\mathcal{G})\\
\quad\quad\quad{\scriptstyle \tau}{\searrow} &  & \!\!\!\!\!\!\!\!
\swarrow{\scriptstyle t}%
\end{array}
\\
H_{\mathrm{loc}}^{1} (\mathfrak{G},\Omega_{\mathrm{loc}}^{0}(\Gamma(E))).
\end{array}
\]

Moreover, as $t$ and $\tau$ are injective, if $\omega\in\Omega_{\mathcal{G}%
,\mathrm{loc}}^{2}(\Gamma(E))$ is closed and $[\underline{\omega
}]=\mathrm{ChW}([\omega])$ then the following conditions are equivalent

\begin{enumerate}
\item[(a)] $[\omega]=0$ on $H_{\mathcal{G},\mathrm{loc}}^{2}(\Gamma(E))$,

\item[(b)] $[\underline{\omega}]=0$ on $H_{\mathrm{loc}}^{2}(\Gamma
(E)/\mathcal{G})$,

\item[(c)] $[\tau(\omega)]=[t(\underline{\omega})]=0$ on $H_{\mathrm{loc}}^{1}
(\mathfrak{G},\Omega_{\mathrm{loc}}^{0}(\Gamma(E)))$.
\end{enumerate}

Hence our definition of $H_{\mathrm{loc}}^{2}(\Gamma(E)/\mathcal{G})$ solves
the problem. It is important to note that if $\omega\in\Omega_{\mathcal{G}%
,\mathrm{loc}}^{2}(\Gamma(E))$ is closed, the form $\underline{\omega}%
\in\Omega^{2}(\Gamma(E)/\mathcal{G})$ determining the class $\mathrm{ChW}%
([\omega])$ could contain non-local terms, as $\underline{\omega}$ depends on
the curvature of a connection $\Theta$ on the principal $\mathcal{G}$-bundle
$\Gamma(E)\rightarrow\Gamma(E)/\mathcal{G}$, and $\Theta$ usually contains
non-local terms. However, the form $t(\underline{\omega})$ obtained by
applying the transgression map $t$ to $\underline{\omega}$ is local.

In this paper we prefer to work with local $\mathcal{G}$-equivariant
cohomology in place of the cohomology of the quotient for several reasons.
Generally, in order to have a well defined quotient manifold, it is necessary
to restrict the group $\mathcal{G}$ to a subgroup acting freely on $\Gamma
(E)$. However, the equivariant cohomology is well defined for arbitrary
actions. Furthermore, the local $\mathcal{G}$-equivariant cohomology can be
related to the cohomology of jet bundles, thus providing new tools for the
study of local anomalies. In terms of local $\mathcal{G}$-equivariant
cohomology the conditions for anomaly cancellation can be expressed in the
following way. Let $\Omega_{\mathcal{G}}^{\mathrm{det\,Ind}D}\in
\Omega_{\mathcal{G}}^{2}(\Gamma(E))$ be the $\mathcal{G}$-equivariant
curvature of the determinant line bundle $\mathrm{det\,Ind}D\rightarrow
\Gamma(E)$ with respect to its natural connection. For free actions we have
$\mathrm{ChW}[\Omega_{\mathcal{G}}^{\mathrm{det\,Ind}D}]= [\Omega
^{\mathrm{det\,Ind}D/\mathcal{G}}]$. Hence, our preceding considerations can
be resumed by saying that if $\Omega_{\mathcal{G}}^{\mathrm{det\,Ind}D}%
\in\Omega_{\mathcal{G},\mathrm{loc}}^{2}(\Gamma(E))$, then  \emph{the local
anomaly is measured by the cohomology class of the } $\mathcal{G}%
$\emph{-equivariant curvature } $\Omega_{\mathcal{G}}^{\mathrm{det\,Ind}D}$
\emph{of the determinant line bundle } $\mathrm{det\,Ind}D\rightarrow
\Gamma(E)$ \emph{on the local } $\mathcal{G}$\emph{-equivariant cohomology }
$H_{\mathcal{G},\mathrm{loc}}^{2}(\Gamma(E))$.

In \cite{localVB} we have shown that, using the variational bicomplex theory,
the local cohomology can be computed in terms of the cohomology of the jet
bundle. By definition, a local functional $\Lambda\in\Omega_{\mathrm{loc}}%
^{0}(\Gamma(E))$ is given by integration over $M$ of a function $L(s,\partial
s)$ depending of the section $s\in\Gamma(E)$ and its derivatives
$\Lambda(s)=\int_{M}L(s,\partial s)\mathrm{vol}_{M}$. The jet bundle
$J^{\infty}(E)$ is the space of Taylor series $j_{x}^{\infty}s$ of sections
$s\in\Gamma(E)$ at points $x\in M$. Hence, the function $L(s,\partial s)$ can
be considered as a function $L\in\Omega^{0}(J^{\infty}(E))$ such that
$(j^{\infty}s)^{\ast}L=L(s,\partial s)$ for every $s\in\Gamma(E)$, and the
Lagrangian density $\lambda=L\mathrm{vol}_{M}\in\Omega^{n}(J^{\infty}(E))$ can
be considered as an $n$-form on $J^{\infty}(E)$. We define a map $\Im
\colon\Omega^{n}(J^{\infty}E)\rightarrow\Omega^{0}(\Gamma(E))$ by setting
$\Im\lbrack\lambda]=\int_{M}(j^{\infty}s)^{\ast}\lambda$ and we have
$\Omega_{\mathrm{loc}}^{0}(\Gamma(E))= \Im(\Omega^{n}(J^{\infty}E))$.

For our study of anomalies we need to consider not only local functionals, but
also local $k$-forms of degree $k>0$. For this reason we extend the map $\Im$
to forms of degree greater than $n$, $\Im\colon\Omega^{n+k}(J^{\infty}E)
\rightarrow\Omega^{k}(\Gamma(E))$ and we set $\Omega_{\mathrm{loc}}^{k}%
(\Gamma(E)) =\Im(\Omega^{n+k}(J^{\infty}E))$. This map can be studied
completely in terms of the jet bundle by means of the variational bicomplex
theory. For $k>1$ the interior Euler operator $I\colon\Omega^{n+k}(J^{\infty
}E) \rightarrow\Omega^{n+k}(J^{\infty}E)$ (a generalization of the
Euler-Lagrange operator) satisfies $I^{2}=I$, $\Im\lbrack\alpha\rbrack
=\Im\lbrack I(\alpha)\rbrack$ and $\Im\lbrack\alpha\rbrack=0$ if and only if
$I(\alpha)=0$, for $\alpha\in\Omega^{n+k}(J^{\infty}E)$. The image of the
interior Euler operator $\mathcal{F}^{k}(J^{\infty}E) =I(\Omega^{n+k}%
(J^{\infty}E))$ is called the space of functional forms, and clearly we have
$\mathcal{F}^{k}(J^{\infty}E)\cong\Omega_{\mathrm{loc}}^{k}(\Gamma(E))$,
$H_{\mathrm{loc}}^{k}(\Gamma(E)) \cong H^{k}(\mathcal{F}^{\bullet}(J^{\infty
}E))$ for $k>0$. Standard results on the variational bicomplex theory can be
used to show that $H^{k}(\mathcal{F}^{\bullet}(J^{\infty}E))\cong
H^{n+k}(J^{\infty}E)$, and in this way the local cohomology is computed in
terms of the cohomology of jet bundles. In a similar way, for the invariant
cohomology, under very general conditions we have $H_{\mathrm{loc}}^{k}%
(\Gamma(E))^{\mathcal{G}}\cong H^{n+k}(J^{\infty}E)^{\mathcal{G}}$ for $k>1$
(see \cite{localVB} for details). Although we do not have a similar result for
equivariant cohomology (see Section \ref{LEC}), we can use these results in
order to study local anomalies in the following way. A necessary condition for
anomaly cancellation is that $\Omega^{\mathrm{det\,Ind}D}$ should be the
exterior differential of a local $\mathcal{G}$-invariant $1$-form. We call
$[\Omega^{\mathrm{det\,Ind}D}] \in H_{\mathrm{loc}}^{2}(\Gamma
(E))^{\mathcal{G}}$ the first obstruction for anomaly cancellation.

We apply these results to gravitational and mixed anomalies in Sections
\ref{metrics} and \ref{mixed} and we show that in these cases the first
obstruction for anomaly cancellation provides necessary and sufficient
conditions for anomaly cancellation.

We conclude that, when the locality conditions are taken into account, the
anomaly cancellation is not related to the topology of $\Gamma(E)/\mathcal{G}$
or $\mathcal{G}$, but to the geometry of the jet bundle.

\section{The transgression maps}

\label{trans}

First we recall some results of equivariant cohomology in the Cartan model
(\emph{e.g. }see \cite{BGV,GS}). We consider a left action of a connected Lie
group $\mathcal{G}$ on a manifold $\mathcal{N}$, \emph{i.e.} a homomorphism
$\rho\colon\mathcal{G}\rightarrow\mathrm{Diff}\mathcal{N}$. We have an induced
Lie algebra homomorphism $\mathfrak{G}\rightarrow\mathfrak{X} (\mathcal{N})$,
$X\mapsto X_{\mathcal{N}}=\left.  \frac{d}{dt}\right|  _{t=0}\rho(\exp(-tX))$.

The space of $\mathcal{G}$-invariant $r$-forms is denoted by $\Omega
^{r}(\mathcal{N})^{\mathcal{G}}$, and the $\mathcal{G}$-invariant cohomology
by $H^{\bullet}(\mathcal{N})^{\mathcal{G}}$. We denote by $\mathcal{P}%
^{k}(\mathfrak{G}, \Omega^{r}(\mathcal{N}))^{\mathcal{G}}$ the space of degree
$k$ $\mathcal{G}$-invariant polynomials on $\mathfrak{G}$ with values in
$\Omega^{r}(\mathcal{N})$. We recall that $\alpha\in\mathcal{P}^{k}
(\mathfrak{G},\Omega^{r}(\mathcal{N}))$ is $\mathcal{G}$-invariant if for
every $X\in\mathfrak{G}$ and every $g\in\mathcal{G}$ we have $\alpha
(\mathrm{Ad}_{g}X)= \rho(g^{-1})^{\ast}(\alpha(X))$. The infinitesimal version
of this condition is
\begin{equation}
L_{Y_{\mathcal{N}}}\left(  \alpha(X)\right)  = k\alpha([Y,X],X,\overset
{(k-1}{\ldots},X)\text{, \quad} \forall X,Y\in\mathfrak{G}.\label{inv}%
\end{equation}
If $\mathcal{G}$ is connected, then condition (\ref{inv}) is equivalent to the
$\mathcal{G}$-invariance of $\alpha$. We assign degree $2k+r$ to the elements
of $\mathcal{P}^{k}(\mathfrak{G}, \Omega^{r}(\mathcal{N}))^{\mathcal{G}}$. The
space of $\mathcal{G}$-equivariant differential $q$-forms is $\Omega
_{\mathcal{G}}^{q} (\mathcal{N})=\bigoplus_{2k+r=q} (\mathcal{P}%
^{k}(\mathfrak{G}, \Omega^{r}(\mathcal{N})))^{\mathcal{G}}$.

The Cartan differential $d_{c}\colon\Omega_{\mathcal{G}}^{q}(\mathcal{N})
\rightarrow\Omega_{\mathcal{G}}^{q+1}(\mathcal{N})$ is defined by
$(d_{c}\alpha)(X)= d(\alpha(X))-\iota_{X_{\mathcal{N}}}\alpha(X)$, and we have
$\left(  d_{c}\right)  ^{2}=0$. The $\mathcal{G}$-equivariant cohomology (in
the Cartan model) of $\mathcal{N}$, $H_{\mathcal{G}}^{\bullet}(\mathcal{N})$,
is the cohomology of the complex $(\Omega_{\mathcal{G}}^{\bullet}%
(\mathcal{N}),d_{c})$.

Let $\omega\in\Omega_{\mathcal{G}}^{2}(\mathcal{N})$ be a $\mathcal{G}%
$-equivariant $2$-form. Then we have $\omega=\omega_{0}+\mu$ where $\omega
_{0}\in\Omega^{2}(\mathcal{N}) ^{\mathcal{G}}$, and $\mu\in\mathrm{Hom}\left(
\mathfrak{G},C^{\infty}(\mathcal{N})\right)  ^{\mathcal{G}}$, \emph{i.e}%
$\emph{.}$\emph{,} $\mu$ is a $\mathcal{G}$-equivariant linear map $\mu
\colon\mathfrak{G}\rightarrow C^{\infty}(\mathcal{N})$. We have $d_{c}%
\omega=0\;$if\ and only if $d\omega_{0}=0$, and $\iota_{X_{\mathcal{N}}}%
\omega_{0}=d(\mu(X)),\;$for every $X\in\mathfrak{G}$. Hence \emph{a closed}
$\mathcal{G}$\emph{-equivariant }$2$\emph{-form is the same as a}
$\mathcal{G}$\emph{-invariant pre-symplectic form and a moment map for it}.

We recall the Berline-Vergne construction of equivariant characteristic
classes (see \cite{BV1,BGV}). Let $\pi\colon P\rightarrow\mathcal{N}$ a
principal $G$-bundle and $\mathcal{G}$ a Lie group acting (on the left) on $P$
by automorphisms. If $A$ is a $\mathcal{G}$\emph{-invariant} connection on $P$
with curvature $F$, we define the equivariant curvature of $A$ by
$F_{\mathcal{G}}(X)= F-A(X_{P})$. Then for every Weil polynomial $f\in
I_{k}^{G}$, the $\mathcal{G}$-equivariant characteristic form associated to
$f$ and $A$ is $f(F_{\mathcal{G}})\in\Omega_{\mathcal{G}}^{2k}(\mathcal{N})$,
It can be seen that $d_{c}(f(F_{\mathcal{G}}))=0$ and that the equivariant
cohomology class $f_{\mathcal{G}}(P)= [f(F_{\mathcal{G}})]\in H_{\mathcal{G}%
}^{2k}(\mathcal{N})$ is independent of the $\mathcal{G}$-invariant connection
$A$.

Finally we recall (e.g. see \cite{BGV}) that if $\mathcal{N}\rightarrow
\mathcal{N}/\mathcal{G}$ is a principal $\mathcal{G}$-bundle we have the
(generalized) Chern-Weil homomorphism $\mathrm{ChW}\colon H_{\mathcal{G}%
}^{\bullet}(\mathcal{N})\rightarrow H^{\bullet}(\mathcal{N}/\mathcal{G})$. If
$A$ is an arbitrary connection on $\mathcal{N}\rightarrow\mathcal{N}%
/\mathcal{G}$ with curvature $F$, and $\alpha\in\Omega_{\mathcal{G}}%
^{q}(\mathcal{N})$, then we have $\mathrm{ChW}([\alpha])=[\mathrm{hor}%
_{A}(\alpha(F))]$, where $\mathrm{hor}_{A}$ is the horizontalization with
respect to the connection $A$. We also use the notation $\underline{\alpha
}=\mathrm{ChW}(\alpha)$. A direct computation shows that we have the following
result, that provides a direct proof of the fact that the Chern-Weil map
$\mathrm{ChW}\colon H_{\mathcal{G}}^{2} (\mathcal{N})\rightarrow H^{2}
(\mathcal{N}/\mathcal{G})$ is an isomorphism.

\begin{proposition}
\label{ChW2}Let $\mathcal{N}\rightarrow\mathcal{N}/\mathcal{G}$ be a principal
$\mathcal{G}$-bundle, and let $A\in\Omega^{1}(\mathcal{N},\mathfrak{G})$ be a
connection form, with curvature $F$. If $\omega=\omega_{0}+\mu\in
\Omega_{\mathcal{G}}^{2}(\mathcal{N})$ is a closed $\mathcal{G}$-equivariant
$2$-form and we define $\alpha\in\Omega^{1}(\mathcal{N})^{\mathcal{G}}$ by
$\alpha=\mu(A)$ then we have $\mathrm{hor}_{A}(\omega(F))=\omega+d_{c}\alpha$.
\end{proposition}

Let us assume that $H^{1}(\mathcal{N})=H^{2}(\mathcal{N})=0$. We denote by
$H^{\bullet}(\mathfrak{G},\Omega^{0}(\mathcal{N}))$ the cohomology of the Lie
algebra $\mathfrak{G}$ with values in $\Omega^{0}(\mathcal{N})$. The following
Proposition can be proved using Formula (\ref{inv})

\begin{proposition}
\label{closed}Let $\omega=\omega_{0}+\mu\in\Omega_{\mathcal{G}}^{2}%
(\mathcal{N})$ be a closed $\mathcal{G}$-equivariant form. If $\rho\in
\Omega^{1}(\mathcal{N})$ satisfies $\omega_{0}=d\rho$, then the map
$\tau_{\rho}\in\Omega^{1}(\mathfrak{G},\Omega^{0}(\mathcal{N}))$ given by
$\tau_{\rho}(X)=\rho(X_{\mathcal{N}})+\mu(X)$ determines a linear map
$\tau\colon H_{\mathcal{G}}^{2}(\mathcal{N})\rightarrow H^{1} (\mathfrak{G}%
,\Omega^{0}(\mathcal{N}))$ which is independent of the form $\rho$ chosen, and
that we call the transgression map $\tau$. If the group $\mathcal{G}$ is
connected, then the transgression map $\tau$ is injective.
\end{proposition}

Now we assume that the action of $\mathcal{G}$ on $\mathcal{N}$ is free, and
$\pi\colon\mathcal{N}\rightarrow$ $\mathcal{N}/\mathcal{G}$ is a principal
$\mathcal{G}$-bundle. Then we can consider the more familiar transgression map
defined as follows

\begin{proposition}
Let $\underline{\omega}\in\Omega^{2}(\mathcal{N}/\mathcal{G})$ be a closed
$2$-form. If $\eta\in\Omega^{1}(\mathcal{N})$ is a form such that $\pi^{\ast
}\underline{\omega}=d\eta$, then the map $t_{\eta}\colon\mathfrak{G}%
\rightarrow\Omega^{0}(\mathcal{N})$, $t_{\eta}(X)=\eta(X_{\mathcal{N}})$
determines a linear map $t\colon H^{2}(\mathcal{N}/\mathcal{G}) \rightarrow
H^{1}(\mathfrak{G},\Omega^{0}(\mathcal{N}))$, which is independent of the form
$\eta$ chosen, and that we call the transgression map $t$. If the group
$\mathcal{G}$ is connected, then the transgression map $t$ is injective.
\end{proposition}

The following Proposition relates the two transgression maps. We use this
result in order to relate our approach to anomalies with the BRST approach.

\begin{proposition}
\label{2T}Let $\omega\in H_{\mathcal{G}}^{2}(\mathcal{N})$ and $\underline
{\omega}= \mathrm{ChW}(\omega)\in H^{2}(\mathcal{N}/\mathcal{G})$. We have
$\tau(\omega)=t(\underline{\omega})$.
\end{proposition}

\begin{proof}
If $\omega=\omega_{0}+\mu$, by Proposition \ref{ChW2} we have $\omega
=\pi^{\ast}\underline{\omega}+d_{c}\alpha$ for some $\alpha\in\Omega
_{\mathcal{G}}^{1}(\mathcal{N})=\Omega^{1}(\mathcal{N})^{\mathcal{G}}$, i.e.
$\omega_{0}=\pi^{\ast}\underline{\omega}+d\alpha$ and $\mu(X)=-\alpha
(X_{\mathcal{N}})$.

Let $\eta\in\Omega^{1}(\mathcal{N})$ be a form such that $\pi^{\ast}%
\underline{\omega}=d\eta$. If we set $\rho=\eta+\alpha$ then $\omega_{0}%
=d\rho$ and for every $X\in\mathrm{Lie}$ $\mathcal{G}$ we have $\tau_{\rho
}(X)=\rho(X_{\mathcal{N}})+\mu(X)=t_{\eta}(X). $
\end{proof}

\section{Local equivariant cohomology\label{LEC}}

Let $p\colon E\rightarrow M$ be a bundle over a compact, oriented $n$-manifold
$M$ without boundary. We denote by $J^{r}E$ its $r$-jet bundle, and by
$J^{\infty}E$ the infinite jet bundle (see \cite{saunders} for the details on
the geometry of $J^{\infty}E$). We recall that the points on $J^{\infty}E$ are
the Taylor series of sections of $E$ and that $\Omega^{k}(J^{\infty}E)=
\underrightarrow{\mathrm{lim}}\Omega^{k}(J^{r}E).$

A diffeomorphism $\phi\in\mathrm{Diff}E$ is said to be projectable if there
exists $\underline{\phi}\in\mathrm{Diff}M$ satisfying $\phi\circ
p=p\circ\underline{\phi}$. We denote by $\mathrm{Proj}E$ the space of
projectable diffeomorphism of $E$, and we denote by $\mathrm{Proj}^{+}E$ the
subgroup of elements such that $\underline{\phi}\in\mathrm{Diff}^{+}M$,
\emph{i.e.} $\underline{\phi}$ is orientation preserving. The space of
projectable vector fields on $E$ is denoted by $\mathrm{proj}E$, and can be
considered as the Lie algebra of $\mathrm{Proj}E$. We denote by $\mathrm{pr}%
\phi$ (resp. $\mathrm{pr}X$) the prolongation of $\phi\in\mathrm{Proj}E$
(resp. $X\in\mathrm{proj}E$) to $J^{\infty}E$.

Let $\Gamma(E)$ be the manifold of global sections of $E$, that we assume to
be not empty. For any $s\in$ $\Gamma(E)$, the tangent space to the manifold
$\Gamma(E)$ is isomorphic to the space of vertical vector fields along $s$,
that is $T_{s}\Gamma(E)\simeq\Gamma(M,s^{\ast}V(E))$.

Let $\mathrm{j}^{\infty}\colon M\times\Gamma(E)\rightarrow J^{\infty}E$,
$\mathrm{j}^{\infty}(x,s)=j_{x}^{\infty}s$ be the evaluation map. We define a
map $\Im\colon\Omega^{n+k}(J^{\infty}E)\longrightarrow\Omega^{k}(\Gamma(E))$,
by $\Im\lbrack\alpha]= \int_{M}\left(  \mathrm{j}^{\infty}\right) ^{\ast
}\alpha$ for $\alpha\in\Omega^{n+k}(J^{\infty}E)$. If $\alpha\in\Omega
^{k}(J^{\infty}E)$ with $k<n$, we set $\Im\lbrack\alpha]=0$. We define the
space of local $k$-forms on $\Gamma(E)$ by $\Omega_{\mathrm{loc}}^{k}%
(\Gamma(E))= \Im(\Omega^{n+k}(J^{\infty}E))\subset\Omega^{k}(\Gamma(E))$. The
local cohomology of $\Gamma(E)$, $H_{\mathrm{loc}}^{\bullet}(\Gamma(E))$, is
the cohomology of $(\Omega_{\mathrm{loc}}^{\bullet}(\Gamma(E)),d)$. The map
$\Im$ induces isomorphisms $H^{k}_{\mathrm{loc}}(\Gamma(E)) \cong H^{n+k}(E)$
for $k>0$ (see \cite{localVB} for details). Note that $\Omega_{\mathrm{loc}%
}^{0}(\Gamma(E))$ is precisely the space of local functions on $\Gamma(E)$.

The group $\mathrm{Proj}E$ acts naturally on $\Gamma(E)$ as follows. If
$\phi\in\mathrm{Proj}E$, we define $\phi_{\Gamma(E)}\in\mathrm{Diff}\Gamma(E)$
by $\phi_{\Gamma(E)}(s)=\phi\circ s\circ\underline{\phi}^{-1}$, for all
$s\in\Gamma(E)$. In a similar way, a projectable vector field $X\in
\mathrm{proj}E$ induces a vector field $X_{\Gamma(E)}\in\mathfrak{X}%
(\Gamma(E))$.

Let $\mathcal{G}$ be a Lie group acting on $E$ by elements $\mathrm{Proj}%
^{+}E$. We define the space of local $\mathcal{G}$-invariant forms
$\Omega_{\mathrm{loc}}^{k}(\Gamma(E))^{\mathcal{G}}$ as the subspace of
$\mathcal{G}$-invariant elements on $\Omega_{\mathrm{loc}}^{k}(\Gamma(E))$,
and the local $\mathcal{G}$-invariant cohomology, $H_{\mathrm{loc}}^{k}%
(\Gamma(E))^{\mathcal{G}}$, as the cohomology of $(\Omega_{\mathrm{loc}%
}^{\bullet}(\Gamma(E))^{\mathcal{G}},d)$. In \cite{localVB} it is shown that
we have $\Omega_{\mathrm{loc}}^{k}(\Gamma(E))^{\mathcal{G}}= \Im(\Omega
^{n+k}(J^{\infty}E)^{\mathcal{G}})$ for $k>0$ and that under certain
conditions $\Im$ induces isomorphisms $H^{k}_{\mathrm{loc}}(\Gamma
(E))^{\mathcal{G}} \cong H^{n+k}(J^{\infty}E)^{\mathcal{G}}$ for $k>1$.

The integration operator extends to a map into equivariant differential forms
(see \cite{equiconn}) $\Im\colon\Omega_{\mathcal{G}}^{n+k} (J^{\infty
}E)\rightarrow\Omega_{\mathcal{G}}^{k}(\Gamma(E))$, by setting $(\Im
\lbrack\alpha])(X)= \Im\lbrack\alpha(X)]$ for every $\alpha\in\Omega
_{\mathcal{G}}^{n+k}(J^{\infty}E)$, $X\in\mathfrak{G}$. The map $\Im$ induces
a homomorphism in equivariant cohomology $\Im\colon H_{\mathcal{G}}%
^{n+k}(J^{\infty}E) \rightarrow H_{\mathcal{G}}^{k}(\Gamma(E))$.

In order to define an adequate notion of local equivariant cohomology we made
the following assumption

\begin{enumerate}
\item[(A1)] We assume that $\mathfrak{G}$ is isomorphic to the space of
sections of a vector bundle $V\rightarrow M$, i.e. $\mathfrak{G}\cong%
\Gamma(V)$. We also assume that the map $\mathfrak{G}\cong\Gamma
(V)\rightarrow\mathrm{proj}E$, $X\mapsto X_{E}$ is a differential operator.
\end{enumerate}

With this assumption, a map $T\colon\bigotimes^{r}\mathfrak{G} \rightarrow
\Omega_{\mathrm{loc}}^{k}(\Gamma(E))$ is said to be local if there exists a
differential operator $t\colon\bigotimes^{r}\mathfrak{G}\rightarrow
\Omega^{n+k} (J^{\infty}E)$ such that $T(X_{1},\ldots,X_{k})= \Im\lbrack
t(X_{1},\ldots,X_{k})]$ for every $X_{1},\ldots,X_{k}\in\mathfrak{G}$. We
denote the space of degree $k$ local polynomials (resp. local $k$-forms) on
$\mathfrak{G}$ with values in $\Omega_{\mathrm{loc}}^{k}(\Gamma(E))$ by
$\mathcal{P}_{\mathrm{loc}}^{r} (\mathfrak{G},\Omega^{k}(\Gamma(E))))$ (resp.
$\Omega_{\mathrm{loc}}^{r} (\mathfrak{G},\Omega_{\mathrm{loc}}^{k}(\Gamma(E))$).

We define the space of local $\mathcal{G}$-equivariant $q$-forms on
$\Gamma(E)$ by $\Omega_{\mathcal{G},\mathrm{loc}}^{q} (\Gamma(E))=\bigoplus
_{2k+r=q} (\mathcal{P}_{\mathrm{loc}}^{k} (\mathfrak{G},\Omega_{\mathrm{loc}%
}^{r} (\Gamma(E))))^{\mathcal{G}}$, and the local $\mathcal{G}$-equivariant
cohomology of \linebreak$\Gamma(E)$, $H_{\mathcal{G},\mathrm{loc}}^{\bullet
}(\Gamma(E))$, as the cohomology of $(\Omega_{\mathcal{G},\mathrm{loc}%
}^{\bullet} (\Gamma(E)),d_{c})$.

\begin{remark}
If a $\mathcal{G}$-equivariant form $\alpha\in\Omega_{\mathcal{G}}%
^{n+k}(J^{\infty}E)$ satisfies that the polynomial map $\alpha\colon
\mathfrak{G}\rightarrow\Omega^{\bullet}(J^{\infty}E)$ is a differential
operator, then $\Im[\alpha] \in\Omega_{\mathcal{G},\mathrm{loc}}^{k}%
(\Gamma(E))$. However, even if we assume that in the definition of the
$\mathcal{G}$-equivariant cohomology of $J^{\infty}E$ we impose that the
polynomial maps $\alpha\colon\mathfrak{G}\rightarrow\Omega^{\bullet}%
(J^{\infty}E)$ are differential operators, $\Im$ will not induce isomorphisms
$H^{n+k}_{\mathcal{G}}(J^{\infty}E) \cong H^{k}_{\mathcal{G},\mathrm{loc}%
}(\Gamma(E))$. For example, if we consider the trivial action of a group
$\mathcal{G}$ on $E$ we have $H^{\bullet}_{\mathcal{G}}(J^{\infty}E) \cong
I^{\mathcal{G}}\otimes H^{\bullet}(J^{\infty}E)$. If $p\in I^{\mathcal{G}}$ is
a Weil polynomial of degree $r$, with $2r>n$, we have by definition $\Im
[p]=0$, and hence the induced map $\Im\colon H_{\mathcal{G}}^{2r} (J^{\infty
}E)\rightarrow H_{\mathcal{G}}^{2r-n}(\Gamma(E))$ is not injective in this case.
\end{remark}

\section{Local anomalies and local equivariant cohomology}

\subsection{Conditions for anomaly cancellation}

Let $E\rightarrow M$ be a fiber bundle, and let $\mathcal{G}$ be a Lie group
acting on $E$ by elements of $\mathrm{Proj}^{+}E$. Let $\{D_{s}:s\in
\Gamma(E)\}$ be a $\mathcal{G}$-equivariant family of elliptic operators
parametrized by $\Gamma(E)$. The determinant line bundle $\mathrm{det\,Ind}%
D\rightarrow\Gamma(E)$ is a $\mathcal{G}$-equivariant line bundle, and is
endowed with a natural $\mathcal{G}$-invariant connection associated to the
Quillen metric. Let $\Omega_{\mathcal{G}}^{\mathrm{det\,Ind}D} \in
\Omega_{\mathcal{G}}^{2}(\Gamma(E))$ be the $\mathcal{G}$-equivariant
curvature of $\mathrm{det\,Ind}D$. We made the following assumption

\begin{enumerate}
\item[(A2)] We assume that $\Omega_{\mathcal{G}}^{\mathrm{det\,Ind}D}$ is a
local $\mathcal{G}$-equivariant form, i.e. that $\Omega_{\mathcal{G}%
}^{\mathrm{det\,Ind}D} \in\Omega_{\mathcal{G},\mathrm{loc}}^{2}(\Gamma(E))$.
\end{enumerate}

In Sections \ref{metrics} and \ref{mixed} we show that for the classical cases
of gravitational and mixed anomalies, assumption (A2) follows form the
Atiyah-Singer Index theorem for families and the results on \cite{equiconn}
and \cite{WP}.

\begin{definition}
We say that the local anomaly corresponding to the $\mathcal{G}$-equivariant
family $\{D_{s}:s\in\Gamma(E)\}$ cancels if the cohomology class of
$\Omega_{\mathcal{G}}^{\mathrm{det\,Ind}D}$ on the local $\mathcal{G}%
$-equivariant cohomology $H_{\mathcal{G},\mathrm{loc}}^{2}(\Gamma(E))$ vanishes.
\end{definition}

\begin{remark}
If the local anomaly cancels, then clearly $c_{1,\mathcal{G}}%
(\mathrm{det\,Ind}D)=0$. However, the converse is not true, as the condition
for anomaly cancellation involves \emph{local} equivariant cohomology.
Furthermore, if the action of $\mathcal{G}$ on $\Gamma(E)$ is free, then we
can consider the quotient bundle $\mathrm{det\,Ind}D/\mathcal{G}%
\rightarrow\Gamma(E)/\mathcal{G}$. Then we have $\mathrm{ChW}([\Omega
_{\mathcal{G}}^{\mathrm{det\,Ind}D}])= c_{1}(\mathrm{det\,Ind}D/\mathcal{G}%
)\in H^{2}(\Gamma(E)/\mathcal{G})$. Hence, if the local anomaly cancels then
we have $c_{1}(\mathrm{det\,Ind}D/\mathcal{G})=0$, but again, this condition
is not sufficient.
\end{remark}

We have $\Omega_{\mathcal{G}}^{\mathrm{det\,Ind}D}= \Omega^{\mathrm{det\,Ind}%
D}+\mu$, where $\mu$ is a moment map for the action of $\mathcal{G}$ on the
pre-symplectic manifold $(\Gamma(E),\Omega^{\mathrm{det\,Ind}D})$. By
definition, the local anomaly cancels if and only if there exists a local
$\mathcal{G}$-invariant $1$-form $\rho\in\Omega_{\mathrm{loc}}^{1}%
(\Gamma(E))^{\mathcal{G}}$ satisfying the conditions $\Omega
^{\mathrm{det\,Ind}D} =d\rho$, and $\mu(X) =-\rho(X_{\Gamma(E)})$, $\forall
X\in\mathfrak{G}$. Hence a necessary condition for the anomaly cancellation is
that $\Omega^{\mathrm{det\,Ind}D}$ should be the exterior differential of a
$\mathcal{G}$-invariant $1$-form. For this reason we made the following

\begin{definition}
\label{conditions}The first obstruction for anomaly cancellation is defined as
the cohomology class $[\Omega^{\mathrm{det\,Ind}D}] \in H_{\mathrm{loc}}%
^{2}(\Gamma(E))^{\mathcal{G}}$ of the curvature of the determinant line bundle
in the local $\mathcal{G}$-invariant cohomology.
\end{definition}

The first obstruction for anomaly cancellation involves local $\mathcal{G}%
$-invariant cohomology, which in \cite{localVB} is shown to be isomorphic to
the cohomology of the $\mathcal{G}$-invariant variational bicomplex. Moreover
under certain conditions we have $H_{\mathrm{loc}}^{2}(\Gamma(E))^{\mathcal{G}%
}\cong H^{n+2}(J^{\infty}E)^{\mathcal{G}}$, and then the first obstruction for
anomaly cancellation can be expressed directly in terms of the jet bundle as
follows. If $\eta\in\Omega^{n+2} (J^{\infty}E)^{\mathcal{G}}$ is a closed form
such that $\Im\lbrack\eta]=\Omega^{\mathrm{det\,Ind}D}$ and the class of
$\eta$ on $H^{n+2}(J^{\infty}E)^{\mathcal{G}}$ does not vanish, then the
anomaly does not cancel. In this way, the techniques developed in
\cite{AndPoh} for computing the invariant cohomology of the variational
bicomplex in terms of Gel'fand-Fuks cohomology can be applied to study the
problem of anomaly cancellation. We apply these results in sections
\ref{metrics} and \ref{mixed} to the case of gravitational and mixed anomalies.

\subsection{Anomaly cancellation and BRST cohomology\label{relationBRST}}

In this section we show that our definition for anomaly cancellation can be
expressed in terms of BRST cohomology. We recall (see \cite{bonora,schmid})
that the BRST cohomology $H_{\mathrm{loc}}^{\bullet} (\mathfrak{G}%
,\Omega_{\mathrm{loc}}^{0}(\Gamma(E)))$ is the Lie algebra local cohomology of
$\mathfrak{G}$ with values in $\Omega_{\mathrm{loc}}^{0}(\Gamma(E))$, that is,
the cohomology of $(\Omega_{\mathrm{loc}}^{\bullet}(\mathfrak{G},
\Omega_{\mathrm{loc}}^{0}(\Gamma(E)),\delta)$. Now we assume that
$H^{2}(\Gamma(E))=H^{1}(\Gamma(E))=0$ and also that $H_{\mathrm{loc}}%
^{2}(\Gamma(E))=H_{\mathrm{loc}}^{1}(\Gamma(E))=0$.

\begin{proposition}
The restriction of the transgression map $\tau$ to $H_{\mathcal{G}%
,\mathrm{loc}}^{2}(\Gamma(E))$ takes values on the BRST cohomology
$H_{\mathrm{loc}}^{1}(\mathfrak{G},\Omega_{\mathrm{loc}}^{0}(\Gamma(E)))$ and
the map \linebreak$\tau\colon H_{\mathcal{G},\mathrm{loc}}^{2}(\Gamma
(E))\rightarrow H_{\mathrm{loc}}^{1} (\mathfrak{G},\Omega_{\mathrm{loc}}%
^{0}(\Gamma(E)))$ is injective for $\mathcal{G}$ connected.
\end{proposition}

\begin{proof}
Let $\omega=\omega_{0}+\mu\in\Omega_{\mathcal{G},\mathrm{loc}}^{2}(\Gamma(E))$
be a closed local $\mathcal{G}$-equivariant $2$-form. As $H_{\mathrm{loc}}%
^{2}(\Gamma(E))=0$, we have $\omega_{0}=d\rho$, for certain $\rho\in
\Omega_{\mathrm{loc}}^{1}(\Gamma(E))$. By the definition of local equivariant
cohomology and assumption (A1) the map $\tau_{\rho}\colon\mathfrak{G}
\rightarrow\Omega_{\mathrm{loc}}^{0}(\Gamma(E))$, $\tau_{\rho}(X)=\rho
(X_{\Gamma(E)})+\mu(X)$ is a local map. The injectiveness of $\tau$ follows
from Proposition \ref{closed}. Note that we can assume that the group is
connected as we are dealing with local anomalies.
\end{proof}

If the action of $\mathcal{G}$\ on $\Gamma(E)$ is free, by Proposition
\ref{2T} we have the following

\begin{proposition}
\label{BRST}Let $\omega\in\Omega_{\mathcal{G},\mathrm{loc}}^{2}(\Gamma(E))$ be
a closed local $\mathcal{G}$-equivariant $2$-form and let $\underline{\omega
}=\mathrm{ChW}(\omega)\in H^{2}(\Gamma(E)/\mathcal{G})$. Then we have
$\tau(\omega)=t(\underline{\omega})$, and in particular $t(\underline{\omega
})\in H_{\mathrm{loc}}^{1} (\mathfrak{G},\Omega_{\mathrm{loc}}^{0}%
(\Gamma(E)))$. Moreover, $t(\underline{\omega})=0$ if and only if the
cohomology class of $\omega$ on $H_{\mathcal{G},\mathrm{loc}}^{2}(\Gamma(E))$ vanishes.
\end{proposition}

With the preceding results, our condition for anomaly cancellation can be
expressed in terms of BRST cohomology in the following way

\begin{theorem}
Let $\{D_{s}:s\in\Gamma(E)\}$ be a $\mathcal{G}$-equivariant family of
elliptic operators satisfying the conditions of assumption \emph{(A2)}. Then
we have $\tau([\Omega_{\mathcal{G}}^{\mathrm{det\,Ind}D}]) \in H_{\mathrm{loc}%
}^{1} (\mathfrak{G},\Omega_{\mathrm{loc}}^{0}(\Gamma(E)))$ and the local
anomaly cancels if and only if $\tau([\Omega_{\mathcal{G}}^{\mathrm{det\,Ind}%
D}])=0$ on the BRST cohomology $H_{\mathrm{loc}}^{1} (\mathfrak{G}%
,\Omega_{\mathrm{loc}}^{0}(\Gamma(E)))$.

In the case of a free action of $\mathcal{G}$ on\ $\Gamma(E)$, we have
$t(c_{1}(\mathrm{det\,Ind}D/\mathcal{G}))= \tau([\Omega_{\mathcal{G}%
}^{\mathrm{det\,Ind}D}])\in H_{\mathrm{loc}}^{1} (\mathfrak{G},\Omega
_{\mathrm{loc}}^{0}(\Gamma(E)))$.
\end{theorem}

\section{Riemannian metrics and gravitational anomalies\label{metrics}}

In this section we apply the preceding considerations to the case of
gravitational anomalies (see \cite{ASZ,freed,Kel}). We consider the family of
Dirac operators ${\not \! \partial}_{g}$ parametrized by the space
$\mathfrak{Met}M$ of Riemannian metrics on $M$, and the action of
diffeomorphisms. First we recall the definition of the equivariant Pontryagin
and Euler forms on the $1$-jet bundle of the bundle of metrics given in
\cite{natconn} and \cite{WP}. Then we show how the equivariant curvature of
the determinant line bundle can be obtained from these constructions on the
jet bundle, and that assumptions (A1) and (A2) hold in this case. Finally, we
use our characterization of local anomaly cancellation in terms of local
equivariant cohomology and the results in \cite{localVB} to obtain necessary
and sufficient conditions for local gravitational anomaly cancellation.

\subsection{Equivariant Pontryagin and Euler forms on $J^{1}\mathcal{M}_{M}$}

Let $M$ be a compact and connected $n$-manifold without boundary, and $TM$ its
tangent bundle. We define its bundle of Riemannian metrics $q\colon
\mathcal{M}_{M}\rightarrow M$ by $\mathcal{M}_{M}=\{g_{x}\in S^{2}(T_{x}%
^{\ast}M):g_{x}$ is positive defined on $T_{x}M\}$. Let $\mathfrak{Met}%
M=\Gamma(M,\mathcal{M}_{M})$ denote the space of Riemannian metrics on $M$. We
denote by $\mathrm{Diff}M$ the diffeomorphisms group of $M$, and by
$\mathrm{Diff}^{+}M$ its subgroup of orientation preserving diffeomorphisms.
We denote by $q_{1}\colon J^{1}\mathcal{M}_{M}\rightarrow M$ the $1$-jet
bundle of $\mathcal{M}_{M}$ and by $\pi\colon FM\rightarrow M$ the linear
frame bundle of $M$. The pull-back bundle $\bar{q}_{1}\colon q_{1}^{\ast
}FM\rightarrow J^{1}\mathcal{M}_{M}$ is a principal $Gl(n,\mathbb{R})$-bundle.

We consider the principal $SO(n)$-bundle $O^{+}M\rightarrow J^{1}
\mathcal{M}_{M}$ where $O^{+}M=\left\{  (j_{x}^{1}g,u_{x})\in q_{1}^{\ast
}FM\colon\text{ }u_{x}\,\text{is\emph{ }}g_{x} \text{-orthonormal and
positively oriented}\right\}  $. In \cite{natconn} it is shown that there
exists a unique connection form $\mbox{\boldmath$\omega$}\in\Omega^{1}%
(O^{+}M,\mathfrak{so}(n))$ (called the universal Levi-Civita connection) on
$O^{+}M$ invariant under the natural action of the group $\mathrm{Diff}^{+}M$.
We denote by $\mbox{\boldmath$\Omega$} $ the curvature form of
$\mbox{\boldmath$\omega$}$.

As the universal Levi-Civita connection $\mbox{\boldmath$\omega$}$ is
$\mathrm{Diff}^{+}M$-invariant, the Berline-Vergne construction of equivariant
characteristic classes (see Section \ref{trans}) can be applied. For any Weil
polynomial $p\in I_{r}^{SO(n)}$ we have the $\mathrm{Diff}^{+}M$-equivariant
characteristic form $p(\mbox{\boldmath$\Omega$}_{\mathrm{Diff}^{+}M})\in
\Omega_{\mathrm{Diff}^{+}M}^{2r}(J^{1}\mathcal{M}_{M})$ corresponding to $p$.
In particular we have the equivariant Pontryagin and Euler forms. If $2r>n$,
by applying the integration map $\Im$ to
$p(\mbox{\boldmath$\Omega$}_{\mathrm{Diff}^{+}M})$, we obtain a closed
$\mathrm{Diff}^{+}M$-equivariant form on $\mathfrak{Met}M$, $\Im\lbrack p(
\mbox{\boldmath$\Omega$} _{\mathrm{Diff}^{+}M})]\in\Omega_{\mathrm{Diff}^{+}%
M}^{2r-n}(\mathfrak{Met}M)$.

Now let us assume that $n=4k-2$ for some integer $k$, and let $p\in
I_{2k}^{SO(n)}$. Then $\omega=\Im\lbrack p(\mbox{\boldmath$\Omega$}
_{\mathrm{Diff}^{+}M})]\in\Omega_{\mathrm{Diff}^{+}M}^{2}(\mathfrak{Met}M)$ is
a closed $\mathrm{Diff}^{+}M$-equivariant $2$-form on $\mathfrak{Met}M$. The
explicit expression of $\omega=\omega_{0}+\mu$ can be found in \cite{WP} where
some geometrical properties of these equivariant $2$-forms are studied. In
particular $\mu\colon\mathfrak{X}(M) \rightarrow\Omega^{0}(\mathfrak{Met}M)$
is given for $g\in\mathfrak{Met}M$ and $X\in\mathfrak{X}(M)$ by $\mu
(X)_{g}=-2k\int_{M}p(\left(  \nabla^{g}X\right)  _{\text{\textsc{A}}}%
,\Omega^{g},\overset{(2k-1}{\ldots\ldots},\Omega^{g}), $ where $\Omega^{g}%
\in\Omega^{2}(M,\mathrm{End}TM)$ is the curvature of the Levi-Civita
connection of $g$, and $\left(  \nabla^{g}X\right)  _{\text{\textsc{A}}}$
denote the skew-symmetric part of $\nabla^{g}X\in\Omega^{0}(M,\mathrm{End}TM)$
with respect to $g$. It follows from this expression of $\mu$ that $\omega
\in\Omega_{\mathrm{Diff}^{+}M,\mathrm{loc}}^{2} (\mathfrak{Met}M)$.

\subsection{Gravitational anomalies}

In this section we apply the preceding considerations to the case of local
gravitational anomalies (see \cite{ASZ,freed,Kel}), and hence we consider the
action of $\mathrm{Diff}^{e}M$, the connected component with the identity on
$\mathrm{Diff}^{+}M$ on the space of Riemannian metrics $\mathfrak{Met}M$. Let
$M$ be a compact spin $n$-manifold, with $n=4k-2$ for some integer $k$, and
let $\rho$ be a representation of $\mathrm{Spin}(n)$. We consider the
$\mathrm{Diff}^{e}M$-equivariant family of chiral Dirac operators
$\{{\not \! \partial}_{g}:g\in\mathfrak{Met}M\}$ coupled to a vector bundle
$V$ associated to the spin frame bundle. The curvature of the determinant line
bundle $\mathrm{det\,Ind\,}{\not \! \partial}\rightarrow\mathfrak{Met}M$ is
given by the Atiyah-Singer index theorem for families in the following way.

Let us consider the principal $SO(n)$-bundle $\mathcal{O}^{+}M\rightarrow
M\times\mathfrak{Met}M$, where $\mathcal{O}^{+}M= \left\{  (u_{x}%
,g)\in\!FM\!\times\!\mathfrak{Met}M\colon u_{x}\,\text{is }g_{x}%
\text{-orthonormal and positively oriented}\right\}  $. The evaluation map
$\mathrm{j}^{1}\colon M\times\mathfrak{Met}M\rightarrow J^{1}\mathcal{M}_{M}$,
admits a lift to the corresponding orthonormal frame bundles $\overline
{\mathrm{j}^{1}}\colon\mathcal{O}^{+}M \rightarrow O^{+}M$, $\overline
{\mathrm{j}^{1}}(u_{x},g)=(u_{x},j_{x}^{1}g)$. The map $\overline
{\mathrm{j}^{1}}$ is a morphism of principal $SO(n)$-bundles and is
$\mathrm{Diff}^{+}M$-equivariant. The pull-back of the universal Levi-Civita
connection $\mbox{\boldmath$\omega$} \in\Omega^{1}(O^{+}M,\mathfrak{so}(n))$
by $\overline{\mathrm{j}^{1}}$ is a $\mathrm{Diff}^{+}M$-invariant connection
form $\hat{\mbox{\boldmath$\omega$}}= \overline{\mathrm{j}^{1}}^{\ast}
\mbox{\boldmath$\omega$} $ on $\mathcal{O}^{+}M$, with curvature
$\hat{\mbox{\boldmath$\Omega$}}= \mathrm{j}^{1\ast}(\mbox{\boldmath$\Omega$})$%
, and $\mathrm{j}^{1\ast} (p_{k}(\mbox{\boldmath$\Omega$} ^{\mathrm{Diff}%
^{+}M}))$ is the $\mathrm{Diff}^{+}M$-equivariant $k$-th Pontryagin form of
$\hat{\mbox{\boldmath$\omega$}}$. By the Atiyah-Singer index theorem for
families we have
\begin{align*}
\Omega_{\mathrm{Diff}^{e}M}^{\mathrm{det\,Ind}{\not \partial }}  &  =\int_{M}
[ \hat{A}(\hat{\mbox{\boldmath$\Omega$}}_{\mathrm{Diff}^{e}M}) \mathrm{ch}%
^{\rho}(\hat{\mbox{\boldmath$\Omega$}}_{\mathrm{Diff}^{e}M})]_{n+2}\\
&  =\Im\lbrack P(\mbox{\boldmath$\Omega$} _{\mathrm{Diff}^{e}M})]\in
\Omega_{\mathrm{Diff}^{e}M,\mathrm{loc}}^{2}(\mathfrak{Met}M)
\end{align*}
where $P=[\hat{A}\mathrm{ch}^{\rho}]_{n/2+1}\in I_{n/2+1}^{O(n)}$ is the
component of $\hat{A}\mathrm{ch}^{\rho}$ of polynomial degree $n/2+1$,
$\hat{A}$ is the $\hat{A}$-genus and $\mathrm{ch}^{\rho}$ denotes the Chern
character with respect the representation $\rho$. Hence the condition of
assumption (A2) is satisfied. That assumption (A1) is also satisfied follows
from the local expression of the lift of $X\in\mathfrak{X}(M)$ to
$\mathcal{M}_{M}$ (see e.g. \cite{natconn}).

\begin{remark}
If we prefer to work with the quotient bundle, we restrict to the subgroup
$\mathrm{Diff}^{0}M$ of diffeomorphisms $\phi\in\mathrm{Diff}M$ such that
$\phi(x_{0})=x_{0}$ and $\phi_{\ast,x_{0}}=\mathrm{id}_{T_{x_{0}}M}$ for
certain $x_{0}\in M$. Then the action of $\mathrm{Diff}^{0}M$ on
$\mathfrak{Met}M$ is free and we have a well defined quotient manifold
$\mathfrak{Met}M/\mathrm{Diff}^{0}M$. The first Chern class of the quotient
bundle is given by $c_{1}\left(  \mathrm{det\,Ind}{\not \! \partial
}/\mathrm{Diff}^{0}M\right)  =\mathrm{ChW}([\Omega_{\mathrm{Diff}^{0}%
M}^{\mathrm{det\,Ind}{\not \partial }}]) \in H^{2}(\mathfrak{Met}%
M/\mathrm{Diff}^{0}M)$. As remarked in the introduction, in this paper we
prefer to work with equivariant cohomology rather than with the cohomology of
the quotient.
\end{remark}

According to Definition \ref{conditions} the first obstruction for anomaly
cancellation is the class $[\Omega^{\mathrm{det\,Ind}{\not \partial }}]\in
H_{\mathrm{loc}}^{2}(\mathfrak{Met}M)^{\mathrm{Diff}^{e}M}$. In \cite{localVB}
it is proved that we have $H_{\mathrm{loc}}^{2}(\mathfrak{Met}%
M)^{\mathrm{Diff}^{e}M} \cong H^{n+2}(J^{\infty}\mathcal{M}_{M}%
)^{\mathrm{Diff}^{e}M}$, and hence, the cohomology class $[\Omega
^{\mathrm{det\,Ind}{\not \partial }}]\in H_{\mathrm{loc}}^{2} (\mathfrak{Met}%
M)^{\mathrm{Diff}^{+}M}$ vanishes if and only if the class of
$P(\mbox{\boldmath$\Omega$})$ on $H^{n+2}(J^{\infty}\mathcal{M}_{M}%
)^{\mathrm{Diff}^{e}M}$ vanishes. We have the following result (see
\cite{localVB})

\begin{theorem}
\label{inj1}The map $I_{k}^{SO(n)}\rightarrow H^{2k}(J^{\infty}\mathcal{M}%
_{M})^{\mathrm{Diff}^{e}M}$, $p\mapsto p(\mbox{\boldmath$\Omega$})$ is
injective for $k\leq n$. Hence a form $p(\mbox{\boldmath$\Omega$})$ is the
exterior differential of a $\mathrm{Diff}^{e}M$-invariant form on $J^{\infty
}\mathcal{M}_{M}$ if and only if $p=0$.
\end{theorem}

Hence, we conclude that \emph{the local gravitational anomaly vanishes if and
only if} $P=0$. Note that the condition for anomaly cancellation is
independent of the manifold $M$ and of the topology of $\mathrm{Diff}^{+}M$ or
$\mathfrak{Met}M/\mathrm{Diff}^{0}M$. It only depends on the dimension $n$ and
the Spin representation $\rho$. This result is in accordance with the
universality character of anomalies expressed in \cite{BCR,BCRS}.

\begin{remark}
The preceding Corollary tell us that if $P\neq0$ it is impossible to find a
\emph{local} counterterm to cancel the anomaly. However, it could be possible
to obtain a non-local counterterm. For example (see \cite{ASZ}), for $M=S^{6}$
we have $c_{1}\left(  \mathrm{det\,Ind}{\not \! \partial}/\mathrm{Diff}%
^{0}M\right) =0$, and hence there exist a non-local counterterm.
\end{remark}

As the space $\mathfrak{Met}M$ is contractible and we have $H_{\mathrm{loc}%
}^{k}(\mathfrak{Met}M) \!\cong\! H^{n+k}(\mathcal{M}_{M}) \cong H^{n+k}(M)=0$
for $k>0$, from Theorems \ref{inj1} and \ref{BRST}  we obtain the following

\begin{corollary}
\label{BRST0}Given $p\in I_{n/2+1}^{SO(n)}$, let $\omega=\omega_{0}+\mu
\in\Omega_{\mathrm{Diff}^{+}M,\mathrm{loc}}^{2}(\mathfrak{Met}M)$ be the
$\mathrm{Diff}^{+}M$-equivariant two form $\omega=\Im\lbrack
p(\mbox{\boldmath$\Omega$}_{\mathrm{Diff}^{+}M})]$. For any $\alpha\in
\Omega_{\mathrm{loc}}^{1}(\mathfrak{Met}M)$ such that $\omega_{0}=d\alpha$,
the cohomology class of $\tau_{\alpha}$ in the local BRST cohomology
$H_{\mathrm{loc}}^{1}(\mathfrak{X}(M),\Omega_{\mathrm{loc}}^{0}(\mathfrak{Met}%
M))$ does not vanish.
\end{corollary}

\section{Connections and mixed Anomalies\label{mixed}}

In this section we made an study of mixed anomalies similar to that of Section
\ref{metrics} for gravitational anomalies. We consider the family of Dirac
operators $\{{\not \! \nabla}_{g,A}\colon g\in\mathfrak{Met}_{M}%
,A\in\mathcal{A}_{P}\}$ parametrized by metrics on $M$ and connections on a
principal bundle $P$, and the action of the group $\mathrm{Aut}P$ of
automorphisms of $P$ (we consider that $\mathrm{Aut}P$ acts on $\mathfrak{Met}%
M$ trough its projection to $\mathrm{Diff}M$). First we recall the definition
of the equivariant characteristic forms on the bundle of connections
introduced on \cite{equiconn}, and using that construction and those in
Section \ref{metrics} we show that assumptions (A1) and (A2) also hold in the
case of mixed anomalies. Finally, we obtain necessary and sufficient
conditions for local mixed anomaly cancellation.

\subsection{The equivariant characteristic forms on the bundle of connections}

We consider a principal $G$-bundle $\pi\colon P\rightarrow M$ over a compact
$n$-manifold $M$. We denote by $\mathcal{A}_{P}$ the space of principal
connections on $P$. Let us recall the definition of the bundle of connections
of $P$ (see \cite{geoconn,Gar2,MS} for details).

Let $\bar{p}\colon J^{1}P\rightarrow P$ be the first jet bundle of $P$. The
action of $G$ on $P$ lifts to an action on $J^{1}P$. We denote by $p\colon
C(P)=J^{1}P/G\rightarrow M=P/G$ the quotient bundle, called the bundle of
connections of $P$. We have a natural identification $\Gamma(C(P))\cong%
\mathcal{A}_{P}$, and we denote by $\sigma_{A}$ the section of $C(P)$
corresponding to $A\in\mathcal{A}_{P}$. The projection ${\bar{\pi}\colon}
J^{1}P\rightarrow C(P)$ is a principal $G$-bundle, isomorphic to the pull-back
bundle $p^{\ast}P\rightarrow C(P)$, that we denote by ${\bar{\pi}\colon
}\mathbb{P}\rightarrow C(P)$. We have the following commutative diagram
\[%
\begin{array}
[c]{ccc}%
\mathbb{P} & \overset{\bar{p}}{\longrightarrow} & P\\
{\scriptstyle\bar{\pi}}\downarrow\text{ \ } &  & \text{ }\downarrow
{\scriptstyle\pi}\\
C(P) & \overset{p}{\longrightarrow} & M
\end{array}
\]
The map $\overline{p}$ is $G$-equivariant, i.e., is a principal $G$-bundle morphism.

The group $\mathrm{Aut}P$ of principal $G$-bundle automorphisms is denoted by
$\mathrm{Aut}P$. If $\phi\in\mathrm{Aut}P$, we denote by $\underline{\phi}%
\in\mathrm{Diff}M$ its projection onto $M$. We denote by $\mathrm{Aut}^{+}P$
the subgroup of elements $\phi\in\mathrm{Aut}P$ such that $\underline{\phi}%
\in\mathrm{Diff}^{+}M$. The kernel of the projection $\mathrm{Aut}%
P\rightarrow\mathrm{Diff}M$ is the gauge group of $P$, denoted by
$\mathrm{Gau}P$.

The Lie algebra of $\mathrm{Aut}P$ can be identified with the space
$\mathrm{aut}P\subset\mathfrak{X}(P)$ of $G$-invariant vector fields on $P$.
The subspace of $G$-invariant vertical vector fields is denoted by
$\mathrm{gau}P$ and can be considered as the Lie algebra of $\mathrm{Gau}P$.
We have an exact sequence of Lie algebras $0\rightarrow\mathrm{gau}P
\rightarrow\mathrm{aut}P\rightarrow\mathfrak{X}(M)\rightarrow0$.

The action of $\mathrm{Aut}P$ on $P$ induces actions on $J^{1}P$ and $C(P)$,
and the maps ${\bar{\pi}}$ and $\bar{p}$ are $\mathrm{Aut}P$-invariant. At the
infinitesimal level, if $X\in\mathrm{aut}P$, we denote by $\underline{X}%
\in\mathfrak{X}(M)$ its projection to $M$, and by $X_{\mathbb{P}}%
\in\mathfrak{X}(\mathbb{P})$, $X_{C(P)}\in\mathfrak{X}(C(P))$ its lift to
$\mathbb{P}=J^{1}P$ and $C(P)$ respectively. That the action of $\mathrm{Aut}%
P$ on $C(P)$ satisfies assumption (A1) follows from the natural identification
$\mathrm{aut}P\cong\Gamma(M,TP/G)$ and the local expression of $X_{C(P)}$
(e.g. see \cite{geoconn}).

The principal $G$-bundle ${\bar{\pi}\colon}\mathbb{P\rightarrow} C(P)$ is
endowed with a canonical $\mathrm{Aut}P$-invariant connection $\mathbb{A}%
\in\Omega^{1}(\mathbb{P},\mathfrak{g})$. This connection can be identified to
the contact form on $J^{1}P$. Alternatively, it can be defined by setting
$\mathbb{A}_{(u,\sigma_{A}(x))}(X)=A_{u}(\bar{p}_{\ast}X)$, for every
connection $A$ on $P$, $x\in M$, $u\in\pi^{-1}(x)$, $X\in T_{(u,\sigma
_{A}(x))}\mathbb{P}$. Let $\mathbb{F}$ be the curvature of $\mathbb{A}$. Again
as $\mathbb{A}$ is $\mathrm{Aut}P$-invariant, we can apply the Berline-Vergne
construction of equivariant characteristic classes. If $f\in I_{k}^{G}$ is a
Weil polynomial of degree $k$ for $G$, we denote by $f(\mathbb{F}%
_{\mathrm{Aut}P})\in\Omega_{\mathrm{Aut}P}^{2k}(C(P))$ the $\mathrm{Aut}%
P$-equivariant characteristic form associated to $f$.

If $2k>n$, by applying the map $\Im$ to $f(\mathbb{F}_{\mathrm{Aut}P})$ we
obtain closed $\mathrm{Aut}^{+}P$-equivariant form on $\mathcal{A}_{P}$. In
particular if $n=2r$ is even and $f\in I_{r+1}^{G}$ then $\omega
=\Im[f(\mathbb{F}_{\mathrm{Aut}P})]\in\Omega_{\mathrm{Aut}^{+}P}%
^{2}(\mathcal{A}_{P})$. We have $\omega=\omega_{0}+\mu$, and the expression of
$\mu\colon\mathrm{aut}P\rightarrow\Omega^{0}(\mathcal{A}_{P})$ is given for
$X\in\mathrm{aut}P$ and $A\in\mathcal{A}_{P}$ by $\mu(X)_{A}=\int_{M}
f(A(X),F_{A},\overset{(r}{\ldots\ldots},F_{A})$, and from this expression we
conclude that $\omega\in\Omega_{\mathrm{Aut}^{+}P,\mathrm{loc}}^{2}%
(\mathcal{A}_{P})$

As usual (see \cite{AS}), we consider the principal $G$-bundle $P\times
\mathcal{A}_{P}\rightarrow M\times\mathcal{A}_{P}$. The evaluation map
$\mathrm{ev}\colon M\times\mathcal{A}_{P}\rightarrow C(P)$, $\mathrm{ev}%
(x,A)=\sigma_{A}(x)$ extends to an $\mathrm{Aut}P$-equivariant map
$\overline{\mathrm{ev}}\colon P \times\mathcal{A}_{P}\rightarrow\mathbb{P}$,
by setting $\overline{\mathrm{ev}}(u_{x},A)=(u_{x},\sigma_{A}(x))$ for every
$x\in M$. Then $\hat{\mathbb{A}}=\overline{\mathrm{ev}}^{\ast}\mathbb{A}$ is a
$\mathrm{Aut}P$-invariant connection on $P\times\mathcal{A}_{P}$, with
curvature $\hat{\mathbb{F}}=\mathrm{ev}^{\ast}\mathbb{F}$, and for every
$f\in\mathcal{I}_{k}^{G}$, $\mathrm{ev}^{\ast}f(\mathbb{F}_{\mathrm{Aut}P})$
is the $\mathrm{Aut}P$-equivariant characteristic form of $\hat{\mathbb{A}}$
associated to $f$.

\subsection{Mixed anomalies}

Now we consider the product bundle $\mathcal{M}_{M}\times_{M}C(P)\!\rightarrow
\! M$. The group $\mathrm{Aut}P$ acts on $C(P)$ as explained above, and acts
on $\mathcal{M}_{M}$ through its projection on $\mathrm{Diff}M$, and hence
$\mathrm{Aut}P$ acts on the product $\mathcal{M}_{M}\times_{M}C(P)$ and on
$J^{\infty}(\mathcal{M}_{M}\times_{M}C(P))$. The two projections $J^{\infty
}(\mathcal{M}_{M}\times_{M}C(P))\rightarrow J^{\infty}\mathcal{M}_{M}$,
$J^{\infty}(\mathcal{M}_{M}\times_{M}C(P))\rightarrow J^{\infty}C(P)$ are
$\mathrm{Aut}P$-equivariant. We denote by the same letter the forms on these
spaces and their pull-backs to $J^{\infty}(\mathcal{M}_{M}\times_{M}C(P))$. In
particular, on $\Omega_{\mathrm{Aut}P}^{\bullet} (J^{\infty}(\mathcal{M}%
_{M}\times_{M}C(P)))$ we have the $\mathrm{Aut}^{+}P$-equivariant Pontryagin
forms $p(\mbox{\boldmath$\Omega$}_{\mathrm{Aut}P})$ coming from $J^{\infty
}\mathcal{M}_{M}$, and the $\mathrm{Aut}P$-equivariant characteristic forms
$f(\mathbb{F}_{\mathrm{Aut}P})$, coming from $J^{\infty}C(P)$.

Let $\beta\colon G\rightarrow\mathrm{Gl}(E)$ be a linear representation of $G$
and let $\mathcal{E}\rightarrow M$ be the vector bundle associated to $P$ and
$\beta$. We denote by $\mathrm{Aut}^{e}P$ the connected component with the
identity in $\mathrm{Aut}P$, and we consider the $\mathrm{Aut}^{e}%
P$-equivariant family of Dirac operators $\{{\not \! \nabla}_{g,A}\colon
g\in\mathfrak{Met}M,A\in\mathcal{A}_{P}\}$. Let us consider the bundle
$\mathcal{Q}=\pi_{1}^{\ast}(P\times\mathcal{A}_{P})\times\pi_{2}^{\ast}
(\mathcal{O}^{+}M)\rightarrow M\times\mathfrak{Met}M\times\mathcal{A}_{P}$,
where $\pi_{1}\colon M\times\mathfrak{Met}M\times\mathcal{A}_{P} \rightarrow
M\times\mathcal{A}_{P}$ and $\pi_{2}\colon M\times\mathfrak{Met}%
M\times\mathcal{A}_{P} \rightarrow M\times\mathfrak{Met}M$ are the
projections. We have the following commutative diagram
\[%
\begin{array}
[c]{ccccc}
&  &  &  & \\
P\times\mathcal{A}_{P} & \overset{\overline{\pi}_{1}}{\longleftarrow} &
\mathcal{Q} & \overset{\overline{\pi}_{1}}{\longrightarrow} & \mathcal{O}%
^{+}(M)\\
\downarrow &  & \downarrow &  & \downarrow\\
M \times\mathcal{A}_{P} & \overset{\pi_{1}}{\longleftarrow} & \!\!\! M \!
\times\! \mathfrak{Met}M \!\times\! \mathcal{A}_{P} \!\!\! & \overset{\pi_{2}%
}{\longrightarrow} & M \times\mathfrak{Met}M
\end{array}
\]

The bundle $\mathcal{Q}$ is a principal $(SO(n) \times G)$-bundle, with
$\mathrm{Aut}^{+}P$-invariant connection $\mathfrak{A}=\overline{\pi}_{1}%
^{*}\hat{\mathbb{A}}+ \overline{\pi}_{2}^{*}\hat{\mbox{\boldmath$\omega$}}$
and curvature $\mathfrak{F}=\pi_{1}^{*}\hat{\mathbb{F}}+ \pi_{2}^{*}%
\hat{\mbox{\boldmath$\Omega$}}$. By the Atiyah-Singer index theorem for
families, the $\mathrm{Aut}^{e}P$-equivariant curvature of the determinant
line bundle is given by%

\begin{align*}
\Omega_{\mathrm{Aut}^{e}P}^{\mathrm{det\,Ind}{\not \nabla }}  & = \int_{M}
\left(  \hat{A}(\mathfrak{F}_{\mathrm{Aut}^{e}P}) \!\wedge\! \mathrm{ch}%
^{\rho} (\mathfrak{F}_{\mathrm{Aut}^{e}P}) \!\wedge\! \mathrm{ch}^{\beta}
(\mathfrak{F}_{\mathrm{Aut}^{e}P}) \right) _{n+2}\\
& = \int_{M}\left( \pi_{2}^{*}\left(  \hat{A}(\hat{\mbox{\boldmath$\Omega$}}%
_{\mathrm{Aut}^{e}P}) \!\wedge\! \mathrm{ch}^{\rho} (\hat
{\mbox{\boldmath$\Omega$}}_{\mathrm{Aut}^{e}P})\right)  \!\wedge\! \pi_{1}%
^{*}\left( \mathrm{ch}^{\beta} (\hat{\mathbb{F}}_{\mathrm{Aut}^{e}P})\right)
\right)  _{n+2}\\
& =\Im\left[  \left(  \hat{A}( \mbox{\boldmath$\Omega$} _{\mathrm{Aut}^{e}P})
\!\wedge\! \mathrm{ch}^{\rho} (\mbox{\boldmath$\Omega$} _{\mathrm{Aut}^{e}%
P})\!\wedge\!\mathrm{ch}^{\beta}(\mathbb{F}_{\mathrm{Aut}^{e}P})\right)
_{n+2}\right] ,
\end{align*}
and hence $\Omega_{\mathrm{Aut}^{e}P}^{\mathrm{det\,Ind}{\not \nabla }}
\in\Omega_{\mathrm{Aut}^{e}P,\mathrm{loc}}^{2} (\mathfrak{Met}_{M}%
\times\mathcal{A}_{P})$ and assumption (A2) is satisfied.

By Definition \ref{conditions} the first obstruction for anomaly cancellation
is
\[
[\Omega^{\mathrm{det\,Ind}{\not \nabla }}]= \Im\left[  \left(  \hat
{A}(\mbox{\boldmath$\Omega$}) \wedge\mathrm{ch}^{\rho}
(\mbox{\boldmath$\Omega$}) \wedge\mathrm{ch}^{\beta} (\mathbb{F})\right)
_{n+2}\right]  \in H_{\mathrm{loc}}^{2} (\mathfrak{Met}M\times\mathcal{A}%
_{P})^{\mathrm{Aut}^{+}P}.
\]
Again (see \cite{localVB}) the map $\Im$ induces an isomorphism
$H_{\mathrm{loc}}^{2}(\mathfrak{Met}M \times\mathcal{A}_{P})^{\mathrm{Aut}%
^{e}P} \cong H^{n+2}(J^{\infty} (\mathcal{M}_{M}\times_{M}C(P)))^{\mathrm{Aut}%
^{e}P}$. Under that isomorphism the first obstruction for anomaly cancellation
corresponds to the cohomology class of the form $\left(  \hat{A}%
(\mbox{\boldmath$\Omega$}) \wedge\mathrm{ch}^{\rho}(\mbox{\boldmath$\Omega$})
\wedge\mathrm{ch}^{\beta} (\mathbb{F})\right)  _{n+2}$ on $H^{n+2}(J^{\infty
}(\mathcal{M}_{M}\times_{M}C(P)))^{\mathrm{Aut}^{e}P}$. We have the following
result (see \cite{localVB})

\begin{theorem}
\label{inj2}The map
\begin{align*}
\bigoplus_{r+s=k}I_{r}^{SO(n)} {\textstyle\bigotimes} I_{s}^{G}  &
\longrightarrow H^{2k}(J^{\infty}(\mathcal{M}_{M}\times_{M}%
C(P)))^{\mathrm{Aut}^{e}P}\\
p\otimes f  &  \mapsto\lbrack p(\mbox{\boldmath$\Omega$})\wedge f(\mathbb{F})]
\end{align*}
is injective for $k\leq n$.
\end{theorem}

Hence, if $Q$ is the component of polynomial degree $n/2+1$ of $\hat
{A}\mathrm{ch}^{\rho}\otimes\mathrm{ch}^{\beta} \in I^{ SO(n) \times G}\cong
I^{SO(n)} \otimes I^{G}$, then the mixed anomaly cancels if and only if $Q=0$.
In particular the gauge and gravitational anomalies cannot cancel between
them. Again the condition for anomaly cancellation does not depend on the
particular manifold $M$ or bundle that we have. It only depends on the
structure group $G$ of $P$ and the dimension $n$ of $M$.

As the space $\mathfrak{Met}M \times\mathcal{A}_{P}$ is contractible and we
have $H_{\mathrm{loc}}^{k}(\mathfrak{Met}M \times\mathcal{A}_{P}) \cong
H^{n+k}( \mathcal{M}_{M} \times_{M} C(P)) \cong H^{n+k}(M)=0$ for $k>0$, by
Theorems \ref{BRST} and \ref{inj1}  we have the following

\begin{corollary}
Let $Q=\sum p_{i} \otimes f_{i} \in I^{SO(n)} \otimes I^{G}$ be a Weil
polynomial of degree $n/2+1$, and let $\omega=\omega_{0}+\mu\in\Omega
_{\mathrm{Aut}^{e}M,\mathrm{loc}}^{2} (\mathfrak{Met}M \times\mathcal{A}_{P})$
be the $\mathrm{Aut}^{e}M$-equivariant two form $\omega=\sum\Im\lbrack
p_{i}(\mbox{\boldmath$\Omega$}_{\mathrm{Aut}^{e}M}) \wedge f_{i}%
(\mathbb{F}_{\mathrm{Aut}^{e}P})\rbrack$. For any $\alpha\in\Omega
_{\mathrm{loc}}^{1} (\mathfrak{Met}M\times\mathcal{A}_{P})$ such that
$\omega_{0}=d\alpha$, the cohomology class of $\tau_{\alpha}$ in the local
BRST cohomology $H_{\mathrm{loc}}^{1}(\mathrm{aut}P,\Omega_{\mathrm{loc}}^{0}
(\mathfrak{Met}M\times\mathcal{A}_{P}))$ does not vanish.
\end{corollary}

\begin{acknowledgement}
This work is supported by Ministerio de Educaci\'{o}n y Ciencia of
Spain, under grant \emph{\#MTM2008-–01386}.
\end{acknowledgement}

\end{document}